\documentclass[usenatbib]{mnras}

\usepackage{graphicx}
\usepackage{hyperref}
\usepackage{xspace}
\usepackage{amsmath}
\usepackage{amssymb}

\def\nmocks{5}

\def\thetamax{\theta_{max}}
\def\mobs{M_*^{\mathrm{(obs)}}}

\def\mstar{M_*}
\def\mstari{M_{*, i}}
\def\mpiv{M_*^{\mathrm{piv}}}

\def\mstarmin{M_{*,{\mathrm{min}}}}
\def\mhalo{M_{h}}
\def\mhaloi{M_{h, i}}

\def\chalo{c_{h}}
\def\chaloi{c_{h, i}}

\def\reff{R_{\mathrm{e}}}

\def\robs{\reff^{\mathrm{(obs)}}}

\def\sigmar{\sigma_{R}}

\def\mstarerr{\sigma_{*, {\mathrm{err}}}}

\def\sigmaeps{\sigma_\epsilon}
\def\epsobs{\epsilon_t^{(\mathrm{obs})}}
\def\epsobsj{\epsilon_{t, j}^{(\mathrm{obs})}}

\def\hyperp{\boldsymbol{\eta}}

\def\individi{\boldsymbol{\psi}_i}
\def\individsamp{\boldsymbol{\psi}_{i}^{(k)}}
\def\data{\mathbf{d}}
\def\datai{\mathbf{d}_i}
\def\wl{\left\{\rm{WL}\right\}}

\def\Sref#1{Section~\ref{#1}\xspace}
\def\Fref#1{Figure~\ref{#1}\xspace}
\def\Tref#1{Table~\ref{#1}\xspace}
\def\Eref#1{Equation~\ref{#1}\xspace}

\def\pr{{\rm P}}

\def\halotools{{\sc Halotools}}

\begin{document}

\title{A Bayesian Hierarchical Approach to Galaxy-Galaxy Lensing}
\author[Sonnenfeld \& Leauthaud]{
Alessandro~Sonnenfeld,$^{1}$\thanks{E-mail:alessandro.sonnenfeld@ipmu.jp}
Alexie~Leauthaud,$^{1, 2}$
\\
$^{1}$Kavli IPMU (WPI), UTIAS, The University of Tokyo, Kashiwa, Chiba 277-8583, Japan \\
$^{2}$ Department of Astronomy and Astrophysics, UCO/Lick Observatory, University of California, 1156 High Street, Santa Cruz, \\ CA 95064, USA\\
}

\maketitle

\begin{abstract}
We present a Bayesian hierarchical inference formalism to study the relation between the properties of dark matter halos and those of their central galaxies using weak gravitational lensing. Unlike traditional methods, this technique does not resort to stacking the weak lensing signal in bins, and thus allows for a more efficient use of the information content in the data. Our method is particularly useful for constraining scaling relations between two or more galaxy properties and dark matter halo mass, and can also be used to constrain the intrinsic scatter in these scaling relations. We show that, if observational scatter is not properly accounted for, the traditional stacking method can produce biased results when exploring correlations between multiple galaxy properties and halo mass. For example, this bias can affect studies of the joint correlation between galaxy mass, halo mass, and galaxy size, or galaxy colour. In contrast, our method easily and efficiently handles the intrinsic and observational scatter in multiple galaxy properties and halo mass. We test our method on mocks with varying degrees of complexity. We find that we can recover the mean halo mass and concentration, each with a $0.1$ dex accuracy, and the intrinsic scatter in halo mass with a $0.05$~dex accuracy. In its current version, our method will be most useful for studying the weak lensing signal around central galaxies in groups and clusters, as well as massive galaxies samples with $\log{\mstar} > 11$, which have low satellite fractions.
\end{abstract}


\section{Introduction}\label{sect:intro}
The relation between the stellar mass of a galaxy and the mass of its host halo is a fundamental ingredient in our understanding of galaxy formation.
At the high mass end of the galaxy population, weak lensing observations have helped constrain this relation up to redshift $z\sim1$ \citep[e.g.][]{Man++06, Lea++12, Cou++15, Z+M15}.
In addition to the well-known stellar-to-halo mass relation (SHMR), the colours of central galaxies are also found to depend on halo mass at fixed stellar mass \citep[see e.g.][]{Hoe++05, Man++06, Tinker2013, Man++16}.
This dependence is interpreted by \citet{Z+M16} as evidence for halo-driven quenching of star formation. Because dark matter halos play an important role in galaxy evolution, we expect additional correlations between the properties of galaxies and those of their host halos. For example, at fixed stellar mass, we might expect correlations between halo mass and galaxy size, age of the stellar population, central black hole mass, to list a few.

Weak gravitational lensing is one of the most direct ways of measuring halos masses, but the signal-to-noise ratio for weak lensing-based halo mass measurements around individual galaxies is typically low.
As a result, it is in general necessary to statistically combine weak lensing measurements around a large number of lenses in order to obtain precise halo mass measurements for population ensembles.

Traditionally, two methods have been used to carry out galaxy-galaxy lensing measurements.
The first approach adopts a maximum-likelihood framework and consists of modeling the contribution from the dark matter halo of each individual lens galaxy to the shear signal, assuming simple parametrized halo density profiles. In order to make the problem computationally tractable, scatter-free scaling relations between galaxy and halo properties are typically assumed \citep[e.g.][]{S+R97, Hud++98, HYG04, LKN05, Han++15}. 

The second, somewhat more popular approach, commonly referred to as ``stacking'',  consists of the derivation of the average weak lensing signal in radial bins around a set of galaxies that have similar observed properties, such as stellar mass \citep[e.g.][]{Hoe++01, Par++05, Man++06b, Lea++12, Vel++14, Bro++16}.
In stacking analyses, the SHMR is constrained by forward modeling the population of galaxies and halos, and comparing it with the stacked weak lensing signal, assuming a parametrized form for the halo density profile.
A significant advantage of stacking over maximum-likelihood methods is the possibility of forward modeling the contribution of satellite galaxies to the weak lensing signal. 

One common limitation to maximum-likelihood and stacking methods, is the need to impose a strict self-similarity relation between galaxies, at some point in the analysis chain.
This in turn results in loss of information, partly limiting the ability of inferring the properties of the SHMR, especially the intrinsic scatter around the mean relation.
In maximum-likelihood studies, the scatter between stellar and halo mass is usually set to zero, and is not inferred directly.
In stacking, when shape measurements around galaxies in a given bin are combined, we are essentially assuming that all objects in the bin are either identical or scaled up versions of one another, depending on the details of the analysis.
Although some information on the intrinsic scatter can be obtained by forward modeling a halo population distribution and comparing it with the stacked signal, the process of stacking erases differences between halos in a given bin, making the resulting data less sensitive to variations in the scatter.

Another limitation of traditional galaxy galaxy lensing methods is the difficulty in exploring dependences of the halo mass on more than one parameter.
For example, let us consider the problem of simultaneously measuring the dependence of halo mass on stellar mass and galaxy size.
If we wish to address this problem with a stacking approach, we need to make bins in the 2-dimensional space defined by stellar mass and size.
This can increase dramatically the number of bins necessary for the analysis, making it difficult to find a number of galaxies sufficient to measure a weak lensing signal in every bin.
Stacking is still effective when the secondary variable can be treated as a binary quantity. 
This is the case, for example, in the study of the relation between halo mass and colour: galaxies can be divided in two colour bins, red and blue, with a clear physical meaning.
However, stacking can be rather cumbersome for studying the dependence of halo mass on variables with a continuous distribution \citep[although see][for a successful 2-dimensional stacked weak lensing analysis]{Z+M17}.

In this work, we present a new method for inferring the distribution of halo masses and its dependence on galaxy properties using weak lensing measurements. The method is based on a Bayesian hierarchical inference formalism. It consists in forward modeling the distribution of galaxies in a multi-dimensional space defined by halo mass, stellar mass, and other properties of interest, and fitting it directly to individual shape measurements.
Similarly to maximum-likelihood methods, we assume scaling relations between galaxy properties, such as stellar mass, and halo mass. 
However, a significant difference in our treatment is that we also allow for the presence of intrinsic and observational scatter, which we infer directly from the data. As we will show later, modeling the scatter is crucial for making accurate inferences in multi-dimensional problems.

A Bayesian hierarchical method for the inference of the mass-concentration relation of galaxy groups and clusters was recently developed and applied to CFHTLenS data by \citet{Lie++17}. 
In this work, we describe how we can use a similar approach to determine the SHMR, its scatter, and to measure the correlation between halo mass and galaxy size at fixed stellar mass.

The goal of this work is to present a method and demonstrate its effectiveness. We do this with a rather simplistic model for the description of the SHMR, but our approach can easily be generalized to the more sophisticated models commonly found in the literature.

This paper is structured as follows. In \Sref{sect:shmr} we briefly review the problem of inferring the SHMR, and its generalization to higher dimensions, highlighting the challenges of this task. 
In \Sref{sect:model} we describe our Bayesian hierarchical inference formalism. 
In \Sref{sect:mock} we describe the mock galaxy catalog and mock weak lensing data used to test our method.
In \Sref{sect:results} we apply our inference method to the mock data, test our ability to infer a dependence of halo mass on galaxy size at fixed stellar mass, and perform various tests to asses the robustness of our analysis. We discuss the results and conclude in \Sref{sect:discuss}.
Throughout our analysis we assume a flat cosmology with $\Omega_M=0.3$, $\Omega_\Lambda=0.7$ and $h=0.7$. Stellar and halo masses are expressed in Solar units. Halo masses are defined as the mass enclosed within a sphere with average density equal to 200 times the critical density of the Universe (usually referred to as $M_{200c}$). The notation $\log{}$ refers to the base 10 logarithm.


\section{The SHMR and its generalization}\label{sect:shmr}

In this Section we introduce the problem of inferring the SHMR with weak lensing data. In particular, we will examine how this problem is typically addressed in stacked weak lensing studies, pointing out possible shortcomings with this approach.

\subsection{Measuring the SHMR}

The SHMR can be described in two, complementary, ways: by specifying the distribution of halo masses at fixed stellar mass, $\pr(\mhalo|\mstar)$, or, vice versa, by describing the distribution in $\mstar$ at fixed $\mhalo$, $\pr(\mstar|\mhalo)$.
The two descriptions are related as follows:
\begin{equation}
\pr(\mstar)\pr(\mhalo|\mstar) = \pr(\mhalo)\pr(\mstar|\mhalo) = \pr(\mstar,\mhalo),
\end{equation}
where $\pr(\mstar)$ is the stellar mass function, $\pr(\mhalo)$ is the halo mass function, and $\pr(\mstar,\mhalo)$ is the joint distribution in the 2-dimensional space defined by stellar and halo mass.
Observationally, when working with a sample selected in stellar mass, $\pr(\mhalo|\mstar)$ is easier to obtain.
However, in the comparison with theoretical models, a description in terms of $\pr(\mstar|\mhalo)$ is usually preferred, because galaxy properties are thought to depend on their dark matter environments and hence $M_h$ is assumed to be the more fundamental property. 
Furthermore, $\pr(\mstar|\mhalo)$ is more stable with respect to the observational scatter on $\mstar$, compared to $\pr(\mhalo|\mstar)$ (we will discuss this point later in this Section).

Stacked weak lensing provides measurements of the radial profile of the excess surface mass density, $\Delta\Sigma$, around galaxies binned in observed stellar mass. 
This is defined as
\begin{equation}
\Delta\Sigma(R) = \bar{\Sigma}(<R) - \Sigma(R), 
\end{equation}
where $\Sigma(R)$ is the surface mass density at projected distance $R$ from a given galaxy, and $\bar{\Sigma}(<R)$ is the average surface mass density within the circle of radius $R$.
The surface mass density around a galaxy is determined by contributions from its stellar component, its dark matter halo, and from neighboring galaxies and halos.

In order to convert stacked measurements into measurements of halo mass, parametrized halo models are usually fit to $\Delta\Sigma$ profiles.
The most straightforward way of interpreting a stacked measurement is by fitting it with a model consisting of a single halo.
This is an appropriate choice when the sample of galaxies consists of central galaxies, and when the radial range of the measurements does not extend far beyond the virial radius of the main halo \citep[see e.g.][]{Man++16}.
The inferred halo mass is close to, but not exactly equal to, the mean mass of the halos in the bin \citep{Man++05}. This is because the excess surface mass density, which is the observable quantity, does not scale linearly with halo mass, and so the mean of the $\Delta\Sigma$ distribution is not equal to the mean of the $\mhalo$ distribution. However, the mean halo mass of the galaxies in the bin can be recovered by applying corrections calibrated with numerical simulations \citep{Man++16}.

With this approach, it is then possible to measure the mean halo mass of galaxies in different stellar mass bins. By making further assumptions on the scatter in halo mass around the mean relation, these measurements can be used to constrain $\pr(\mhalo|\mstar)$.
However, one caveat is that bins are, by necessity, defined in terms of the {\em observed} stellar mass. Due to Eddington bias, the mean true stellar mass of galaxies in a bin is in general lower than the mean of the observed values. 
The inferred SHMR is then a relation between observed stellar mass and true halo mass. Although this distinction might seem pedantic, it can have important implications, especially when looking for secondary correlations between halo mass and galaxy properties at fixed stellar mass, as we will show later.

A more complex approach consists of forward modeling the stacked weak lensing signal created by a population of galaxies and halos with a halo occupation distribution (HOD) model \citep[e.g.][]{Lea++12, Tin++13, Z+M15}.
Given a theoretically motivated halo mass function and a functional form for $\pr(\mstar|\mhalo)$, one can predict the stellar mass function $\pr(\mstar)$ and the stacked weak lensing signal in different bins.
Comparing these predictions with the observed stellar mass function and the measured weak lensing signal allows to constrain the form of the SHMR.
Through such a procedure, it is possible to account for the effect of the environment and for the presence of satellite galaxies. 
Moreover, forward modeling constrains the scatter around the SHMR, as changing the scatter modifies the distribution of halo masses within each stellar mass bin. However, this method needs to use information on the stellar mass function as an additional constraint that is independent from the weak lensing. Hence, in order to apply this method, it is necessary to work with complete samples, or at least, samples that are complete within redshift slices \citep{Z+M15}.

In all practical applications of HOD forward model methods, observational errors in $M_*$ measurements are, more or less explicitly, treated as an additional source of scatter.
$\pr(\mstar | \mhalo)$ is typically described as a Gaussian in $\log{\mstar}$ with dispersion $\sigma_{\log{\mstar}}$. Both intrinsic scatter and observational scatter contribute to this dispersion. 
In principle, the two sources of scatter can be deconvolved. In practice, this is usually not done carefully, for simplicity and due to difficulties estimating accurate stellar mass uncertainties \citep[see discussion in subsection 4.2 of][]{Lea++12}. 
As a result, the inferred SHMR is, also in this case, an observed stellar-to-halo mass relation (OSHMR), $\pr(\mobs | \mhalo)$, which is the convolution between the distribution in true stellar mass and observational errors:
\begin{equation}
\pr(\mobs | \mhalo) \propto \int d\mstar \pr(\mobs | \mstar) \pr(\mstar | \mhalo) 
\end{equation}
If the intrinsic scatter is Gaussian in $\log{\mstar}$ and observational errors on $\log{\mstar}$ are drawn from a Gaussian with zero mean and scatter $\sigma_{*,err}$ (i.e. $\pr(\mobs | \mstar)$ is a Gaussian in $\log{\mobs}$ centred on $\log{\mstar}$ and with dispersion $\sigma_{*,err}$), then $\pr(\mobs | \mhalo)$ is itself a Gaussian with the same mean as $\pr(\mstar | \mhalo)$ and with variance given by
\begin{equation}
\sigma_{\log{\mstar}}^2 = \sigma_{*,int}^2 + \sigma_{*,err}^2,
\end{equation}
where $\sigma_{*,int}$ is the intrinsic scatter. 
This means that the inferred scatter is the quadrature sum of the intrinsic and the observational scatter, and that the inferred mean stellar mass at fixed halo mass is unbiased.

Although intrinsic and observational scatter affect the OSHMR in the same way, the physical interpretation of the two sources of scatter is drastically different: the former is set by processes determining the buildup of stellar mass in dark matter halos, the latter is purely artificial and goes to zero in the limit of perfect observations.

\subsection{Correlations with two or more variables}\label{ssec:introbias}

As we are about to show, the distinction between observed and intrinsic quantities is crucial when measuring correlations between halo mass and two or more galaxy properties simultaneously. 
Let us consider the problem of determining the dependence of halo mass on galaxy size, at fixed stellar mass. 
Naively, one could proceed as follows: 1) select a sample of galaxies in a bin of stellar mass, 2) sub-divide this sample according to whether a galaxy lies above or below the mass-size relation, 3) measure the stacked weak lensing signals for the two samples with different sizes, and 4) compare the inferred surface mass density profile.

Such an approach produces a biased result. The reason is that, observationally, we have no access to the true stellar mass of a galaxy, but only to a noisy estimate of it. At fixed {\em observed} stellar mass, objects with smaller size are more likely to be intrinsically less massive galaxies that have scattered in the bin due to observational uncertainty. These galaxies will on average live in less massive halos. Therefore, such a measurement would appear to show a correlation between size and halo mass even if such a correlation is not present.

\begin{figure}
 \includegraphics[width=\columnwidth]{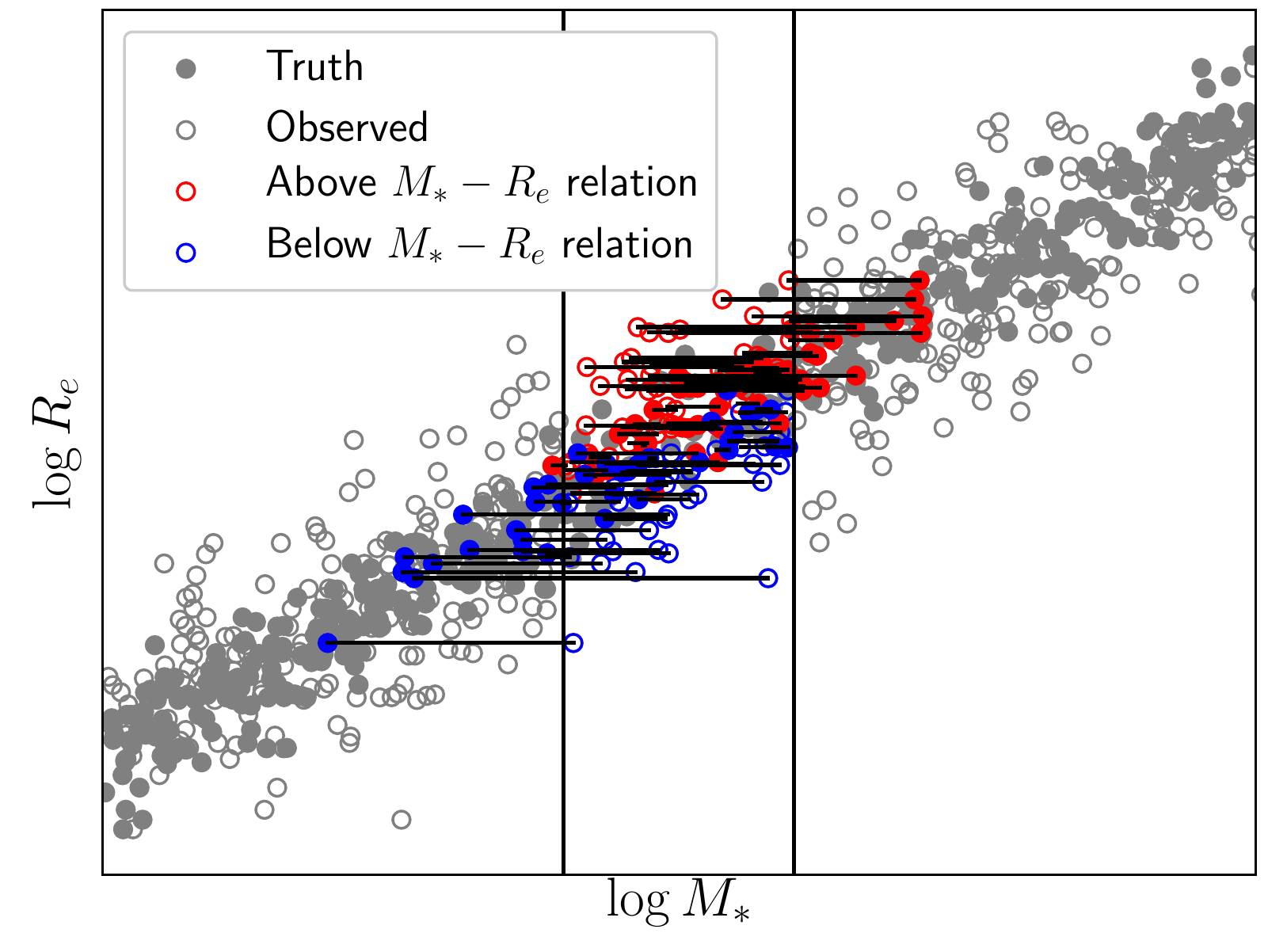}
 \caption{Illustration of the bias introduced when splitting a sample of galaxies by size at fixed observed stellar mass. The plot shows stellar mass vs. effective radius for a mock sample of galaxies. True values are shown as filled circles, while observed values, generated by adding a Gaussian scatter in the logarithm of the stellar mass, are marked by empty circles. The two vertical lines mark the bounds of a bin in observed stellar mass. 
Objects marked in blue (red) are galaxies that belong to the bin in observed stellar mass, and that have a smaller (larger) size compared to the average mass-size relation for galaxies of the same observed stellar mass. 
Horizontal lines connect true and observed stellar masses for the objects in the observed stellar mass bin. 
Most of the objects scattered into the bin from lower (higher) stellar masses are marked in blue (red), indicating that they have smaller (higher) sizes compared to the average for their observed stellar mass. 
To enhance the effect of the bias, we have assumed, when making this mock, that sizes are proportional to the square of the stellar mass, a correlation four times stronger than observed for massive early-type galaxies. 
}
 \label{fig:toy}
\end{figure}

A toy representation of the problem is illustrated in \Fref{fig:toy}, where we plot the effect of observational scatter in stellar mass on a mock mass-size relation. 
We make a bin in observed stellar mass, and divide it in two subsamples, based on whether the observed data points lie above or below the mass-size relation. 
As \Fref{fig:toy} shows, objects scatter into the bin from higher and lower true stellar masses. 
Most of the objects entering the bin from the lower mass side have smaller sizes than the average (blue circles), while most of the objects scattered from the higher stellar mass region have larger sizes (red circles). 
As a result, the average true stellar mass of the larger size subsample is actually larger than that of the smaller size subsample, even though the two subsamples share the same observed stellar mass.

The importance of this effect is higher the stronger the correlation between mass and size, and the larger the observational scatter in stellar mass. 
For the example in \Fref{fig:toy}, we have artificially steepened the mass-size relation for a better visualization of the bias.
Nevertheless,
in order to robustly determine a correlation between size (or any other quantity) and halo mass at fixed stellar mass, the underlying correlation between mass and size and the effects of observational uncertainties must be properly modeled and taken into account.
This bias does not pose a fundamental problem to stacking: in fact, it can be avoided by forward modeling the effects of observational scatter, making a clear distinction between true and observed quantities. However, this bias is also naturally avoided with the inference method presented in the next Section.


\section{The model}\label{sect:model}

In this Section we describe the statistical inference formalism we use for our analysis (in subsection \ref{ssec:form}), the model used to describe the relation  between galaxies and their halos (subsection \ref{ssec:shmr}) and the model adopted to fit weak lensing data (subsection \ref{ssec:like}).
Subsection \ref{ssec:impsamp} will be dedicated to technical aspects of the computations required to perform the fit.
For the sake of clarity, we begin by describing these for a simplified version of the problem, adding more complexity step by step.

\subsection{The inference problem}\label{ssec:form}

Let us consider a sample of galaxies selected in observed stellar mass. 
We would like to infer the distribution of host halo masses, the intrinsic stellar masses of the sample, and the relation between the two quantities. This is equivalent to determining the distribution of galaxies in the 2-dimensional space defined by true stellar and halo mass. 
Let us assume that this distribution can be described analytically by a set of parameters $\hyperp$. We call these the {\em hyper-parameters} of the model and refer to the distribution of stellar and halo masses, given by the hyper-parameters, as $\pr(\mstar, \mhalo | \hyperp)$. 
We wish to infer the posterior probability distribution of the hyper-parameters given weak lensing and stellar mass data $\data$, $\pr(\hyperp|\data)$.
From Bayes theorem,
\begin{equation}\label{eq:bayes1}
\pr(\hyperp|\data) \propto \pr(\hyperp)\pr(\data|\hyperp),
\end{equation}
where $\pr(\hyperp)$ is the prior probability distribution on the hyper-parameters and $\pr(\data|\hyperp)$ is the likelihood of observing the data given the model.
The latter term expands as follows:
\begin{multline}\label{eq:nasty}
\pr(\data|\hyperp) = \idotsint dM_{*,1} dM_{h,1}\ldots dM_{*,N} dM_{h,N} \times \\
\pr(\data|M_{*,1}, M_{h,1}, \ldots, M_{*,N}, M_{h,N}) \times \\
\pr(M_{*,1}, M_{h,1}, \ldots, M_{*,N}, M_{h,N}|\hyperp).
\end{multline}
In order to calculate the likelihood of observing the data given the hyper-parameters, we need to marginalize over all possible values of the individual stellar and halo masses, resulting in a $2\times N$-dimensional integral, where $N$ is the number of lenses in the sample.
The data consists of stellar mass measurements of all lenses and weak lensing shape measurements on background sources.
In principle, each background source is lensed by all lenses, therefore the above integral cannot be simplified without further assumptions.

In maximum-likelihood weak lensing methods, the problem is simplified by assuming no scatter in the SHMR: the term $\pr(M_{*,1}, M_{h,1}, \ldots, M_{*,N}, M_{h,N}|\hyperp)$ reduces to a product of delta functions, and \Eref{eq:nasty} results in a trivial integral.
Here we wish to model the scatter, therefore we make a different assumption:
we assume that lenses are isolated from each other, and that each background source is only lensed by one lens galaxy.
The likelihood term then simplifies to
\begin{equation}\label{eq:factorize}
\pr(\data|\hyperp) = \prod_i \pr(\datai|\hyperp),
\end{equation}
with
\begin{equation}\label{eq:integral1}
\pr(\datai|\hyperp) = \iint dM_{*,i} dM_{h,i} \pr(\datai|\mstari, \mhaloi)\pr(\mstari, \mhaloi|\hyperp).
\end{equation}
Here $\datai$ indicates the data relative to the $i$-th lens only. This consists of a stellar mass measurement of the lens galaxy, and shape measurements of background sources located around the lens. \Sref{sect:results} will describe how source galaxies are assigned to lenses. In short, we will only consider the lensing effect on sources located roughly within the expected virial radius of each lens, because the assumption of isolated lenses will break down at large distances where the effects of neighboring halos become significant. The effects of the isolated lens assumption will be tested in subsection \ref{ssec:full}.

\Eref{eq:integral1} relates the distribution of stellar and halo masses $\pr(\mstari, \mhaloi | \hyperp)$, which we wish to infer, to the data.
The first term in the integral in the right hand side of \Eref{eq:integral1} is the likelihood of observing the data $\datai$ given the values of the stellar and halo mass of the $i$-th galaxy. To calculate this term, we must specify a model describing the galaxy+halo system and predict the expected values of the shear and stellar mass measurements given $\mstar$ and $\mhalo$.
Our choice for the model halo density profile will be discussed in \ref{ssec:like}.

\Eref{eq:integral1} shows the hierarchical nature of the problem: the likelihood of the data is given by the values of parameters describing individual objects, $\mstari$ and $\mhaloi$. The probability distribution for these parameters $\pr(\mstari, \mhaloi | \hyperp)$ is in turn specified by the hyper-parameters $\hyperp$. $\pr(\mstari, \mhaloi | \hyperp)$ can be thought of as a prior on stellar and halo mass, where the parameters describing this prior are free, and have priors of their own.

Provided that our model allows us to compute the likelihood term $\pr(\datai | \mstari, \mhaloi)$, we can then use Equations \ref{eq:bayes1} and \ref{eq:factorize} to calculate the posterior probability distribution of the hyper-parameters, through the integrals given by \Eref{eq:integral1}.

The Bayesian hierarchical inference formalism described above is very general and can be applied to any problem in which {\em i)} measurements are carried out on a family of objects that can be described by a finite number of parameters and {\em ii)} these parameters can be described as being drawn from a single distribution for the whole sample.
From here on we will make assumptions specific to the weak lensing problem at hand.

\subsection{The stellar-to-halo mass relation}\label{ssec:shmr}

We now describe the distribution in halo mass at fixed stellar mass, $\pr(\mhalo|\mstar)$. To do so, we assume the following form for the distribution in stellar and halo masses:
\begin{equation}
\pr(\mstari, \mhaloi|\hyperp) = \mathcal{S}(\mstari|\hyperp)\mathcal{H}(\mhaloi|\hyperp, \mstari).
\end{equation}
We describe the halo mass term $\mathcal{H}(\mhalo | \hyperp, \mstar)$ as a Gaussian distribution 
\begin{equation}\label{eq:mhalodist}
\mathcal{H}(\mhalo | \hyperp, \mstar) = \frac{1}{\sqrt{2\pi}\sigma_h}\exp{\left\{-\frac{(\log{\mhalo} - \mu_h(\mstar))^2}{2\sigma_h^2}\right\}},
\end{equation}
with a mean scaling with stellar mass as
\begin{equation}\label{eq:muhalo}
\mu_h(\mstar) = \mu_{h,0} + \beta_h(\log{\mstar} - \log{\mpiv})
\end{equation}
and dispersion $\sigma_h$, with $\mpiv$ being an arbitrary pivot point that can be left constant in the analysis.

\Eref{eq:muhalo} corresponds to a power-law relation between stellar and halo mass.
We make this assumption to simplify the presentation of the method, but the model can be trivially generalized to more complex descriptions of the distribution of halo masses at fixed stellar mass commonly found in the literature.
In subsection \ref{ssec:brokenpl} we will consider a modified version of this model, in which the relation between stellar and halo mass is treated as a broken power-law. 

We model the stellar mass term, $\mathcal{S}$, as a skewed Gaussian distribution in the logarithm of the stellar mass,
\begin{equation}\label{eq:fullskew}
\mathcal{S}(\mstar|\hyperp) = \frac{1}{\sqrt{2\pi}\sigma_*}\exp{\left\{-\frac{(\log{\mstar} - \mu_*)^2}{2\sigma_*^2}\right\}}\Phi(\log{M_*}),
\end{equation}
with
\begin{equation}\label{eq:skew}
\Phi(\log{\mstar}) = 1 + \mathrm{erf}\left(\alpha_*\frac{\log{\mstar} - \mu_*}{\sqrt{2}\sigma_*}\right).
\end{equation}
The choice of a skewed Gaussian is motivated by the need to model a sample selected in stellar mass.
For example, for a sample obtained by selecting galaxies above a given value of observed stellar mass, $\mstarmin$, we expect the distribution in true stellar mass to have a sharp (yet continuous, due to observational errors) drop for $\mstar < \mstarmin$, which can be captured by large values of the parameter $\alpha_*$ in \Eref{eq:skew}.
The hyper-parameters introduced so far are
\begin{equation}
\hyperp \equiv \left\{\mu_{h,0},\sigma_h,\beta_h,\mu_*,\sigma_*,\alpha_* \right\}.
\end{equation}
A brief description of each parameter is provided in \Tref{tab:hyperp}.

\subsection{The likelihood term and halo profile}\label{ssec:like}

As discussed in Subsection \ref{ssec:form}, in order to infer the posterior probability distribution of the hyper-parameters we need to calculate the likelihood of the data given the values of stellar and halo mass of each object.
For each lens galaxy, the data consists of an observed stellar mass $\mobs$ and weak lensing shape measurements $\wl$.
The likelihood can then be separated as
\begin{equation}\label{eq:liketerms}
\pr(\data | \mstar, \mhalo) = \pr(\mobs | \mstar) \pr(\wl | \mstar, \mhalo).
\end{equation}
Assuming known Gaussian uncertainties on $\log{\mstar}$, $\mstarerr$, the stellar mass term in the right-hand side becomes
\begin{equation}
\pr(\mobs | \mstar) = \frac{A}{\sqrt{2\pi}\mstarerr}\exp{\left\{ - \frac{(\log{\mstar} - \log{\mobs})^2}{2\mstarerr^2}\right\}},
\end{equation}
where $A$ is a normalization constant such as the integral over all possible values of the observed stellar mass, set by the lower limit $\mstarmin$, is unity:
\begin{equation}\label{eq:norm1}
\int_{\mstarmin}^{\infty} d \log{\mobs} \pr(\mobs | \mstar) = 1.
\end{equation}
In other terms, \Eref{eq:norm1} states that the probability of a lens galaxy having an observed value of $M_*$ between $\mstarmin$ and infinity, given the fact that it is part of the selected sample, is one.

Calculating the weak lensing term of the likelihood requires making assumptions on the density profile of the dark matter halo. 
{\em We assume a spherical Navarro Frenk and White profile for the dark matter distribution} \citep[NFW,][]{NFW97}:
\begin{equation}\label{eq:nfw}
\rho(r) = \frac{\rho_0}{r/r_s(1 + r/r_s)^2}.
\end{equation}
An NFW profile can be fully described by two parameters. We choose halo mass and concentration, defined as the ratio between the virial radius and the scale radius $r_s$. Consistently with our definition of halo mass, $\mhalo = M_{200c}$, we define the virial radius as the radius of the sphere enclosing an average density equal to 200 times the critical density of the Universe, $r_{200c}$.
The concentration is then
\begin{equation}
\chalo \equiv \frac{r_{200c}}{r_s}.
\end{equation}
We assume the following mass-concentration relation for the halos:
\begin{equation}\label{eq:cmrel}
\pr(\chalo | \mhalo) = \frac{1}{\sqrt{2\pi}\sigma_c}\exp{\left\{-\frac{(\log{\chalo} - \mu_c(\mhalo))^2}{2\sigma_c^2}\right\}},
\end{equation}
with
\begin{equation}\label{eq:maccio}
\mu_c(\mhalo) = \mu_{c, 0} + \beta_c(\log{\mhalo} - \log{\mhalo}^{\mathrm{piv}}).
\end{equation}
The parameters $\mu_{c,0}$, $\beta_c$ and the scatter $\sigma_c$ are to be inferred from the data, while $\mhalo^{\mathrm{piv}}$ is an arbitrary pivot point.

Having introduced an additional parameter, the concentration, the likelihood term (\Eref{eq:integral1}) becomes
\begin{multline}\label{eq:integral2}
\pr(\datai | \hyperp) = \iiint d\log{\mstari} d\log{\mhaloi} d\log{\chaloi} \times \\
\pr(\datai | \mstari, \mhaloi, \chaloi) \pr(\chaloi | \mhaloi) \pr(\mstari, \mhaloi | \hyperp)
\end{multline}
Note that, since we expressed all our probability distributions in terms of the logarithm of the parameters $\mstar$, $\mhalo$ and $\chalo$, the integration variables have been changed accordingly.

The contribution from the baryons in the lens galaxy to the weak lensing signal is modeled with a circular de Vaucouleurs profile \citep{deV48}, with half-light radius inferred from the data.
We also assume that the measurements of the stellar mass, $\mobs$, provide an {\em accurate} estimate of the baryonic mass.
This will not necessarily be the case in practical applications of this method: typical stellar mass measurements are based on a set of assumptions on the stellar population, which could bias the inference. Particularly important is the assumption of a stellar initial mass function (IMF): if the true IMF is different from the one assumed to obtain $\mobs$, then the stellar mass estimate will be biased.
For simplicity, we ignore possible variations of the stellar IMF in this theoretical study. As we will show, the contribution of the stellar mass to the lensing signal is in any case very small. 
Finally, we assume that the redshift of the lens galaxy is known exactly, and that the dark matter halo is centred at the galaxy position.

Given our mass model, we can calculate the predicted reduced-shear signal of the lens at a given position $\boldsymbol\theta$ and source redshift $z_s$, $g(\boldsymbol\theta, z_s)$, 
defined as
\begin{equation}\label{eq:redshear}
g(\boldsymbol\theta, z_s) = \frac{\gamma(\boldsymbol\theta, z_s)}{1 - \kappa(\boldsymbol\theta, z_s)}.
\end{equation}
$\kappa(\boldsymbol\theta, z_s)$ and $\gamma(\boldsymbol, z_s)$ are respectively the dimensionless surface mass density and the complex shear generated by the lens at image position $\boldsymbol\theta$ for a source at redshift $z_s$.
We refer to \citet{Bar96, W+B2000} for details on the lensing properties of NFW halos.

The reduced-shear is a complex quantity. The tangential component of it, $g_t$, is equal to the tangential ellipticity induced by the lens on a circular source at $z_s$. 
The likelihood of an individual shape measurement $\epsobs$ from a source at {\em known} redshift $z_s$ given the model is
\begin{equation}\label{eq:onesourcelike}
\pr(\epsobs | \mstar, \mhalo, \chalo) = \frac{1}{\sqrt{2\pi}{\sigmaeps}}\exp{\left\{-\frac{(g(\boldsymbol\theta, z_s) - \epsobs)^2}{2\sigmaeps^2}\right\}},
\end{equation}
where $\sigmaeps$ is the observational uncertainty on the tangential shear, typically dominated by the intrinsic dispersion in galaxy shapes. 

In cases when the estimate of the source redshift is noisy, i.e. most practical applications of weak lensing, the uncertainty on the source redshift, $\Delta z_s$, introduces an additional source of error on the tangential shear.
This error is due to the dependence of the reduced shear $g$ on the lensing critical surface mass density $\Sigma_{cr}$, and, in the limit of small $\Delta z_s$, is given by
\begin{equation}\label{eq:zerr}
\left|\frac{\partial g}{\partial z_s}\right| \Delta z_s = \left|\frac{\partial g}{\partial \Sigma_{cr}}\right| \left|\frac{\partial \Sigma_{cr}}{\partial z_s}\right| \Delta z_s = \frac{|\gamma|}{(1-\kappa)^2} \left|\frac{\partial \ln{\Sigma_{cr}}}{\partial z_s}\right| \Delta z_s.
\end{equation}
Under the assumption that the uncertainties on the source shape and redshift are independent, the term above can be added in quadrature to $\sigmaeps$ to obtain the final uncertainty on the observed tangential shear.

This additional source of uncertainty is typically much smaller than the intrinsic shape noise, since it is on the order of the magnitude of the shear.
However, \Eref{eq:zerr} is an approximation: since the critical density is not linear in $z_s$, we cannot simply propagate an uncertainty on the redshift into one in $\Sigma_{cr}$. In principle, the likelihood term \ref{eq:onesourcelike} should be calculated by obtaining the model tangential shear for all possible values of the source redshift and marginalizing over $z_s$. Ignoring the non-linearity of $\Sigma_{cr}$ with respect to $z_s$ can in principle introduce a systematic bias. However, we will show that this bias is very small. 

The likelihood of the weak lensing measurements given the halo mass, concentration, and stellar mass, is given by the following product over all shape measurements:
\begin{equation}
\pr(\wl | \mhalo, \chalo, \mstar) = \prod_j \pr(\epsobsj | \mstar, \mhalo, \chalo) 
\end{equation}

\begin{figure}
 \includegraphics[width=\columnwidth]{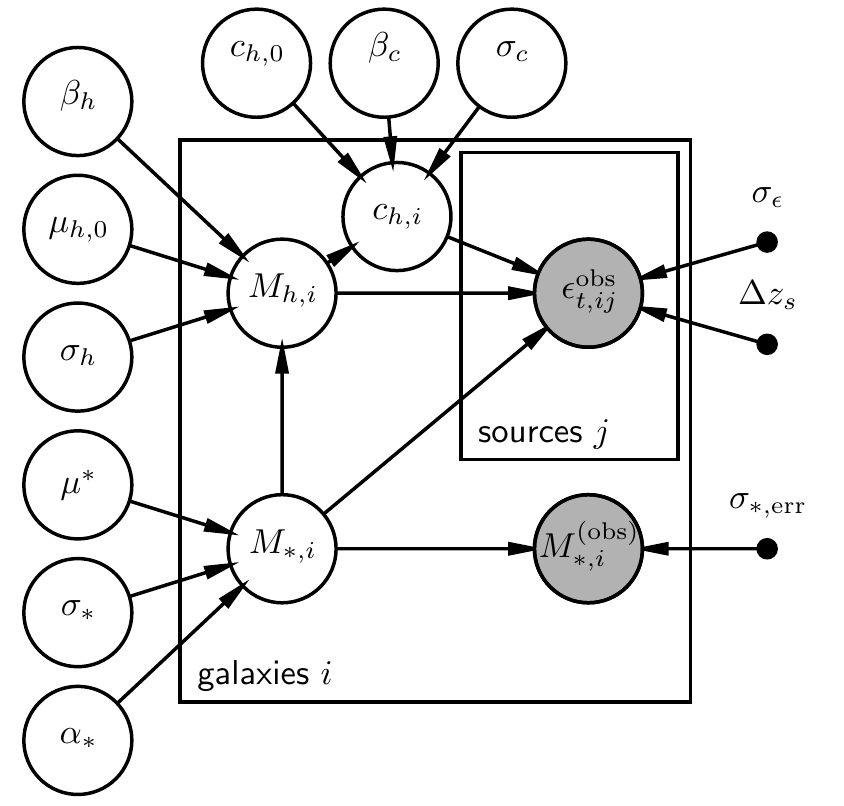}
 \caption{Probabilistic graphical model of the SHMR Bayesian hierarchical inference problem with weak lensing data described in Subsections \ref{ssec:form} through \ref{ssec:like}. 
Circles indicate probability distributions, while dots indicate fixed quantities. Shaded regions refer to observed quantities, while unshaded ones are free parameters. Arrows indicate the dependence between the parameters.
Tables indicate ensembles of parameters.
The hyper-parameters, shown as unshaded circles on the left-hand side of the figure, determine the distributions for the individual stellar and halo masses. Halo masses determine halo concentrations, together with the hyper-parameters describing the mass-concentration relation. Halo mass, concentration and stellar mass determine the tangential ellipticity of background sources.
This is a simplified version of the full problem treated in this paper.}
 \label{fig:pgm1}
\end{figure}

At this point we have described all ingredients necessary for the inference of the hyper-parameters of the model in the context of the assumptions made so far. In \Fref{fig:pgm1} we display a probabilistic graphical model of the inference problem, showing its hierarchical structure. Hyper-parameters, represented as circles on the upper and left-hand side of the figure, feed into lower level variables, the individual stellar and halo mass and the concentration, which in turn determine the observed quantities: the observed stellar mass and source shapes.

\subsection{Sampling the posterior probability distribution}\label{ssec:impsamp}

We would like to measure the posterior probability distribution of the hyper-parameters using a Markov Chain Monte Carlo (MCMC).
In order to evaluate the posterior at each step of the chain we must calculate the three-dimensional integral in \Eref{eq:integral2} for each lens galaxy. This is a computationally expensive task because the likelihood is not an analytical function of halo mass, concentration and stellar mass.
Following \citet{Sch++15} we use Monte Carlo integration together with importance sampling to evaluate these integrals: we sample the likelihood for each object beforehand and then use these samples to do the marginalization over the individual parameters for each drawn value of the hyper-parameters, as explained below.

To simplify the notation, we refer to the individual lens parameters collectively as $\individi \equiv \{\mstari, \mhaloi, \chaloi\}$. 
We introduce a prior on the distribution in $\individi$. This prior, which we call {\em interim prior} and label $I$, should be broad enough to cover the region of parameter space where the likelihood is nonzero (our choice for the interim prior will be discussed later).
We consider the posterior probability distribution of $\individi$ given the data and the interim prior, $\pr(\individi | \datai, I)$. This can be written as, 
\begin{equation}
\pr(\individi | \datai, I) \propto \pr(\individi | I) \pr(\datai | \individi).
\end{equation}
Using the above we can write \Eref{eq:integral2} as
\begin{equation}
\pr(\datai | \hyperp) = \int d\individi 
\frac{\pr(\individi | \datai, I)}{\pr(\individi | I)} \pr(\individi | \hyperp)
\end{equation}
up to a multiplicative constant irrelevant for the problem.
At this point we draw samples $\{\individsamp\}$ of sufficiently large size $N$ from $\pr(\individi | \datai, I)$ with an MCMC and approximate the above integral with the sum over the samples
\begin{equation}\label{eq:impsamp}
\pr(\datai | \hyperp) \approx \frac{1}{N} \sum_k \frac{1}{\pr(\individsamp | I)} \pr(\individsamp | \hyperp). 
\end{equation}
The advantages of this approximation are that $\pr(\individi | \datai)$ only needs to be sampled once at the beginning of the inference, and that the triple integral reduces to a sum.
We choose a uniform distribution in the range $(10, 13)$ as a prior on $\log{\mstari}$, a Gaussian distribution with mean $13.0$ and dispersion $0.5$ on $\log{\mhaloi}$, and a Gaussian distribution with mean $0.8$ and dispersion $0.3$ on $\log{\chalo}$. The exact choice of the interim prior does not affect the results of the inference, as it is divided out in \Eref{eq:impsamp}.

\subsection{The full problem}\label{ssec:fullmodel}

We can now generalize the method to include a secondary dependence of halo mass on a property of its central galaxy. We choose galaxy size to be this parameter.
We introduce the effective radius variable $\reff$ and update the model to account for the dependence of halo and stellar mass on $\reff$.
We assume that stellar mass, halo mass and effective radius are now drawn from a distribution $\pr(\mstar, \mhalo, \reff | \hyperp)$ with the following form:
\begin{equation}
\pr(\mstar, \mhalo, \reff | \hyperp) = \mathcal{S}(\mstar | \hyperp) \mathcal{R}(\reff | \mstar, \hyperp) \mathcal{H}(\mhalo | \mstar, \reff, \hyperp).
\end{equation}
Here $\mathcal{S}(\mstar | \hyperp)$ is the same skewed Gaussian as in \Eref{eq:fullskew}. $\mathcal{R}(\reff | \mstar, \hyperp)$ is the following Gaussian
\begin{equation}\label{eq:reterm}
\mathcal{R}(\reff | \mstar, \hyperp) = \frac{1}{\sqrt{2\pi}\sigmar}\exp{\left\{-\frac{(\log{\reff} - \mu_R(\mstar))^2}{2\sigmar^2}\right\}}
\end{equation}
with mean
\begin{equation}\label{eq:remu}
\mu_R(\mstar) = \mu_{R, 0} + \beta_R(\log{\mstar} - \log{\mpiv})
\end{equation}
and dispersion $\sigmar$. $\mathcal{H}(\mhalo | \mstar, \reff, \hyperp)$ has the same form as the Gaussian in \Eref{eq:mhalodist}, but its mean is updated as follows to include a dependence on stellar mass density
\begin{equation}\label{eq:fullmuhalo}
\mu_h(\mstar, \reff) = \mu_{h,0} + \beta_h(\log{\mstar} - \log{\mpiv}) + \xi_h\log{(\Sigma_*/\Sigma_{*,0})},
\end{equation}
where $\Sigma_* = \mstar/(2\pi\reff^2)$ and $\log{\Sigma_{*,0}}= \log{\mpiv} - \log{2\pi} - 2\mu_{R,0}$ (the stellar mass density of a galaxy with $\mstar=\mpiv$ and average size for its mass).
We choose to parametrize the model with a dependence of halo mass on stellar mass density rather then directly on effective radius to minimize correlations between the model parameters. Stellar mass and size are correlated, and it is difficult with weak lensing data alone to determine which of the two is the fundamental parameter on which halo mass depends. On the other hand, since $\reff$ scales with a power of $\mstar$ close to $0.5$, massive galaxies occupy roughly a region of constant stellar mass density as a function of mass. Using $\mstar$ and $\Sigma_*$ in \Eref{eq:fullmuhalo} allows us to decouple the dependence on mass from that on size, helping in the interpretation of the results.
The full list of hyper-parameters is summarized in \Tref{tab:hyperp}, among with a short description of each parameter.
\begin{table*}
 \caption{Hyper-parameters of the model. The horizontal line divides the hyper-parameters between those relative to the simplified model, described in subsections \ref{ssec:shmr} and \ref{ssec:like}, and the ones appearing only in the full version of the model, introduced in subsection \ref{ssec:fullmodel}.}
 \label{tab:hyperp}
 \begin{tabular}{cclc}
 \hline
& Hyper-parameter & Description & Prior \\
 \hline
& $\mu_{h,0}$ & Mean $\log{\mhalo}$ at $\log{\mstar} = 11.2$ & Uniform$(11, 15)$\\
& $\sigma_h$ & Intrinsic scatter in $\log{\mhalo}$ & Uniform$(0, 2)$ \\
& $\beta_h$ & Power-law index of $\mhalo$-$\mstar$ correlation & Uniform$(-3, 3)$ \\
Simplified model & $\mu_*$ & Mean of Gaussian component of stellar mass distribution & Uniform$(10, 12)$\\
(subsection \ref{ssec:shmr}) & $\sigma_*$ & Dispersion of Gaussian component of stellar mass distribution & Uniform$(0, 2)$\\
& $\alpha_*$ & Skewness parameter of stellar mass distribution & Uniform in $\log$ $(-1, 1)$\\
& $\mu_{c,0}$ & Mean $\log{\chalo}$ at $\log{\mhalo} = 13$ & Uniform$(0, 2)$ \\
& $\sigma_c$ & Intrinsic scatter in $\log{\chalo}$ & Uniform$(0, 1)$ \\ 
& $\beta_c$ & Power-law index of $\chalo$-$\mhalo$ correlation & Uniform$(-1, 1)$ \\
\hline
& $\xi_h$ & Power-law index of $\mhalo$-$\Sigma_*$ correlation & Uniform$(-2, 2)$\\
(Full model only) & $\mu_{R,0}$ & Mean $\log{\reff}$ at $\log{\mstar} = 11.2$ & Uniform$(-1, 2)$ \\
& $\sigma_R$ & Intrinsic scatter in $\log{\reff}$ & Uniform$(0, 1)$ \\
& $\beta_R$ & Power-law index of $\reff$-$\mstar$ correlation & Uniform$(-1, 2)$ \\
 \hline
 \end{tabular}
\end{table*}

\begin{figure}
 \includegraphics[width=\columnwidth]{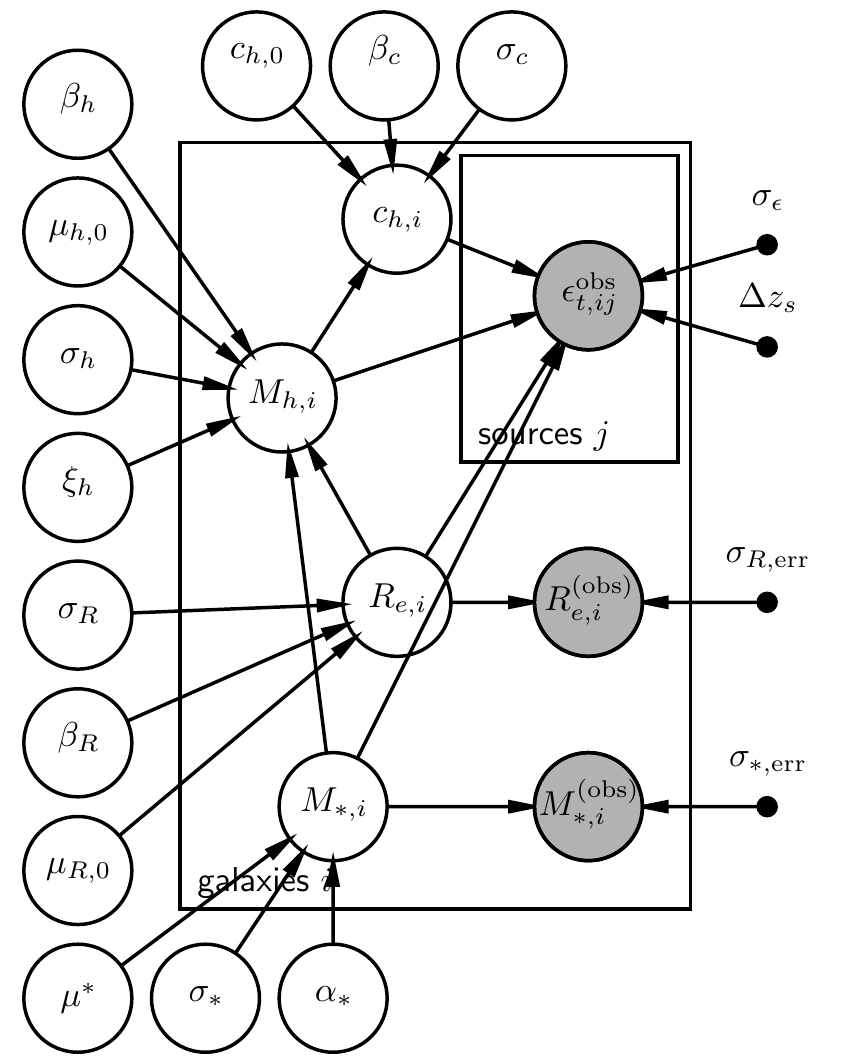}
 \caption{Probabilistic graphical model of the SHMR Bayesian hierarchical inference problem described in subsection \ref{ssec:fullmodel}.
}
 \label{fig:pgm2}
\end{figure}

The likelihood must also be updated to account for the observed effective radius. \Eref{eq:liketerms} becomes
\begin{multline}
\pr(\data | \mstar, \mhalo, \chalo, \reff) = \pr(\mobs | \mstar) \pr(\robs | \reff) \times \\
\pr(\wl | \mstar, \reff, \mhalo, \chalo).
\end{multline}
Here we have assumed that measurements of stellar mass and effective radius are independent. This, however, is not a critical assumption.
A probabilistic graphical model of the problem is shown in \Fref{fig:pgm2} .

\subsection{Possible model extensions}

In the model considered so far, the average halo mass scales with stellar mass and stellar mass density following a power-law relation.
Observational studies of the SHMR, however, indicate that the relation between stellar and halo mass is not a single power-law at all mass scales: the dependence of $\mhalo$ on $\mstar$ is steeper at the high mass end, compared to the low mass regime, the transition occurring around $\log{\mstar}\sim10.5$ \citep[see e.g.][]{Lea++12}.
Our model can be modified to allow for an SHMR with different slopes in different mass ranges: we will show one such example in subsection \ref{ssec:brokenpl}, where we describe the relation between stellar and halo mass as a broken power-law.

Another implicit assumption in our model is that the SHMR does not depend evolve with time. This approximation is reasonable if the sample of lenses covers a thin slice in redshift, but breaks down over a broad redshift range \citep[see e.g.][]{Mos++10}. Nevertheless, we can in principle account for evolution by adding a redshift dependence term to the expression for the average halo mass, \Eref{eq:fullmuhalo}.
In fact, our method can allow us to measure the time evolution of the SHMR without resorting to dividing the sample of lens galaxies in redshift bins.

Finally, we have not discussed the possibility of a dependence on galaxy type of the SHMR. Current constraints from weak lensing \citep{Man++16} and satellite kinematics \citep{Mor++11} suggest that blue central galaxies live in less massive halos compared to red centrals of the same stellar mass (although see \Sref{sect:discuss} for a possible issue with measuring such a correlation with binning and stacking).
While a dependence of the SHMR on galaxy type can be included in our model, we assume for simplicity a single SHMR for all galaxies. Strictly speaking, then, the results of the tests presented in this work will only be relevant for situations in which the sample of lens galaxies is constructed by selecting only objects of a given type.


\section{Mock observations}\label{sect:mock}

We follow a semi-empirical approach to create $\nmocks$ different sets of mock weak lensing observations.
For each set, we use different ingredients to create the population of lens galaxies. We generate mocks of varying complexity to test the impact that different approximations have on the inference. In the next two subsections we describe the most simple and the most complex models among these mock lens populations. We further create an additional intermediate set of three mocks that will be described in \Sref{sect:results} -- these will be used in \Sref{sect:results}. 
\Tref{tab:mocks} summarizes the characteristics of each mock sample.
\begin{table}
 \caption{Mock lens samples. The second column indicates which form of the SHMR has been used. Columns 3 to 5 indicate whether miscentering, the presence of satellite galaxies and the effects of neighboring halos (the 2-halo term) are included in the model.}
 \label{tab:mocks}
 \begin{tabular}{cccccc}
 \hline
 Label & SHMR & Miscent. & Sat. & 2-halo\\
 \hline
A & Power-law & N & N & N \\
B & Power-law & Y & N & N \\
C & B13 & Y & N & N \\
D & B13 & Y & Y & N \\
E & B13 & Y & Y & Y \\
 \hline
 \end{tabular}
\end{table}

\subsection{A toy model (model A)}\label{ssec:mockA}

The most simple model we consider, labeled ``mock A", is generated by drawing stellar and halo masses from 
 from a very similar distribution to the one assumed in subsection \ref{ssec:shmr}. 
We take a Gaussian distribution in $\log{\mstar}$ with mean $11.0$ and dispersion $0.4$. We draw a large number of objects from this distribution, apply a $0.15$~dex observational uncertainty to the stellar mass, apply a cut selecting only objects with $\log{\mobs} > 11$, and finally draw a subsample of 5,000 objects. We then generate halo masses from a Gaussian distribution, with a mean that scales as a power of the (true) stellar mass according to \Eref{eq:muhalo}, with parameters $\mu_{h,0} = 13$, $\beta_h=1.5$, $\log{\mpiv}=11.2$, and with dispersion $\sigma_h=0.4$. Essentially, we are using the same form as \Eref{eq:mhalodist} to describe the SHMR.

We distribute the galaxies in redshift using a Gaussian distribution centred at $z=0.2$, with dispersion $0.1$ and truncated at a minimum redshift $z_{\mathrm{min}}=0.1$ to ensure that each galaxy contributes with an appreciable lensing signal.
We assign a size to each galaxy, drawn from a mass-size relation of the same form as \Eref{eq:reterm} and \Eref{eq:remu}, with $\mu_{R,0}=0.80$, $\beta_R=0.57$ and $\sigma_R=0.16$, as measured by \citet{New++12} using SDSS data, corrected to a Chabrier IMF.
Note that we do not assume any additional dependence of size on halo mass.

We model the mass distribution of each system with a spherical NFW profile describing the halo and a circularly symmetric de Vaucouleurs profile describing the stars. 
To set the scale radius of each halo, we use a mass-concentration relation of the form given by \Eref{eq:cmrel}, with $\mu_{c,0} = 0.72$, $\beta_c = -0.098$, and $\sigma_c = 0.1$, as suggested by dark matter only simulations \citep[see e.g.][]{Mac++08}.
Finally, we assume that each galaxy is exactly at the center of its dark matter halo, ignoring the possibility of miscentering, and that each halo is at a virtually infinite projected distance from any other halo, neglecting the contribution from neighboring halos to the lensing signal.

\subsection{The most complex model (model E)}\label{ssec:mockE}

For the most complex of our mock lens realizations, we make use of the package \halotools\ \citep{Hea++16}.
We draw a sample of galaxies and host halos using the \citet[][B10 from here on]{Beh++10} SHMR and halo catalogs from the Bolshoi simulation at $z=0$. 
We take half of the simulation box, preserving the size along line-of-sight direction, then, as for mock A, we apply a $0.15$~dex observational uncertainty to the stellar masses, and apply a stellar mass cut by selecting only objects with $\log{\mobs} > 11$.
The final sample results in $\sim4,500$ galaxies, 16\% of which are satellites.
Galaxy sizes are generated in the same way as mock A.

With this simulation we wish to test the effects of the environment on our inference, therefore we drop the approximation of galaxies being at infinite distance from each other, and use the full halo position information from the Bolshoi simulation.
However, for simplicity, we still model halos and galaxies with analytical density profiles, similarly to mock A.
In order for the simulation to be realistic, we need halos to have a finite mass.
We then model the mass distribution of each halo with a smoothly truncated spherical NFW profile, following \citet{BMO09}:
\begin{equation}\label{eq:BMO}
\rho(r) = \frac{\rho_0 r_s}{r/r_s (1 + r/r_s)^2}\left(\frac{r_t^2}{r^2 + r_t^2}\right)^2.
\end{equation}
We set the truncation radius $r_t$ to be the same as the value of $r_{200c}$ for each halo, 
and assume the same mass-concentration relation used for mock A.
The distribution of stellar mass is modeled as a de Vaucouleurs profile.

In addition to the halos of the galaxies in the sample, we model the contribution to the lensing signal of all halos with $M_h>10^{12.5}M_\odot$.
We allow for miscentering between galaxy and halos. This is added as a random shift in projection between the centers of the two mass components, drawn from a Gaussian distribution with zero mean and $10\,\rm{kpc}$ dispersion.
In the literature, the term miscentering is sometimes used in a broader sense, to indicate situations in which the galaxy at the center of the halo is not the most massive among those bound to the main halo. The B10 model used for this mock already includes such cases.

We treat the sample of galaxies and halos as a thin screen of lenses at $z=0.3$.
The Bolshoi box is $250\,\rm{Mpc}/h$ on the side. For our fiducial cosmology, this corresponds to an area of 243 square degrees at $z=0.3$. For the sake of computational time, we only use half of this area, leaving us with a $\sim120$ square degrees simulation. This is similar to the area covered by the first-year shear catalog of the Hyper Suprime-Cam (HSC) survey \citep{Aih++17, Man++17}.

The absence of a lens redshift distribution is the only element that is more simple compared to mock A.
Otherwise, this mock contains a variety of features that are present in the real Universe and could bias our inference method: satellite galaxies, the effects of neighboring halos, and miscentering.


\subsection{Mock source sample}\label{ssec:wlmock}

We generate a mock source sample with a uniform distribution in the image plane of $20\,\rm{arcmin}^{-2}$ and a Gaussian redshift distribution centred at $z=1.0$ and with dispersion $\sigma_z=0.5$.
The two components of the complex ellipticity of each source are drawn from a Gaussian centred in zero and with dispersion $\sigma_{\epsilon} = 0.27$.
We then add lensing distortion.
For each source, we calculate the reduced-shear resulting from the contribution of all halos within a physical projected distance in the image plane smaller than ten times their virial radius. In mock E, all lenses are at the same redshift and the reduced-shear can be calculated using \Eref{eq:redshear}, where the surface mass density $\kappa$ and the shear $\gamma$ are just the sum of the individual values of $\kappa$ and $\gamma$ of each contributing lens.
Although we also consider mocks with a distribution in lens redshifts, in all such cases lenses are assumed to be at infinite distance from each other, so that \Eref{eq:redshear} can be used directly, with $\kappa$ and $\gamma$ being the surface mass density and complex shear of the only lens affecting the source.

Given the value of the reduced shear $g$, and given a value for the intrinsic complex ellipticity of the source, $\epsilon_s$,
the observed complex ellipticity is given by \citep{S+S97}
\begin{equation}
\epsilon = \left\{\begin{array}{ll} \dfrac{\epsilon - g}{1 - g^*\epsilon} & \rm{for}\,|g| < 1 \\
& \\
\dfrac{1 - g\epsilon^*}{\epsilon^* - g^8} & \rm{for}\,|g| > 1\end{array}\right. .
\end{equation}
Finally, we add a photo-z measurement error of $\Delta z_s = 0.1$.

The parameters used to simulate weak lensing measurements are set to resemble those of the HSC survey \citep{Man++17, Tan++17}.


\section{Results}\label{sect:results}

\subsection{Mock A: source photo-z uncertainty}

We fit the model introduced in \ref{ssec:fullmodel} to the data from mock A (weak lensing, stellar masses, and effective radii, see subsections \ref{ssec:mockA} and \ref{ssec:wlmock}). We only use sources located within a cone of angular radius $\thetamax$ centred on each lens. This corresponds to $300\,\rm{kpc}$ in projected physical distance at the redshift of our lenses.
This choice approximates what we would do when dealing with actual data, in order to stay in a regime where our isolated lens approximation is valid. 
Although no effects from the environment are included in mock A, we still apply this cut in projected source distance to check whether we can still make a meaningful inference on the SHMR with weak lensing data not extending too far from the lens.

\begin{figure*}
 \includegraphics[width=\textwidth]{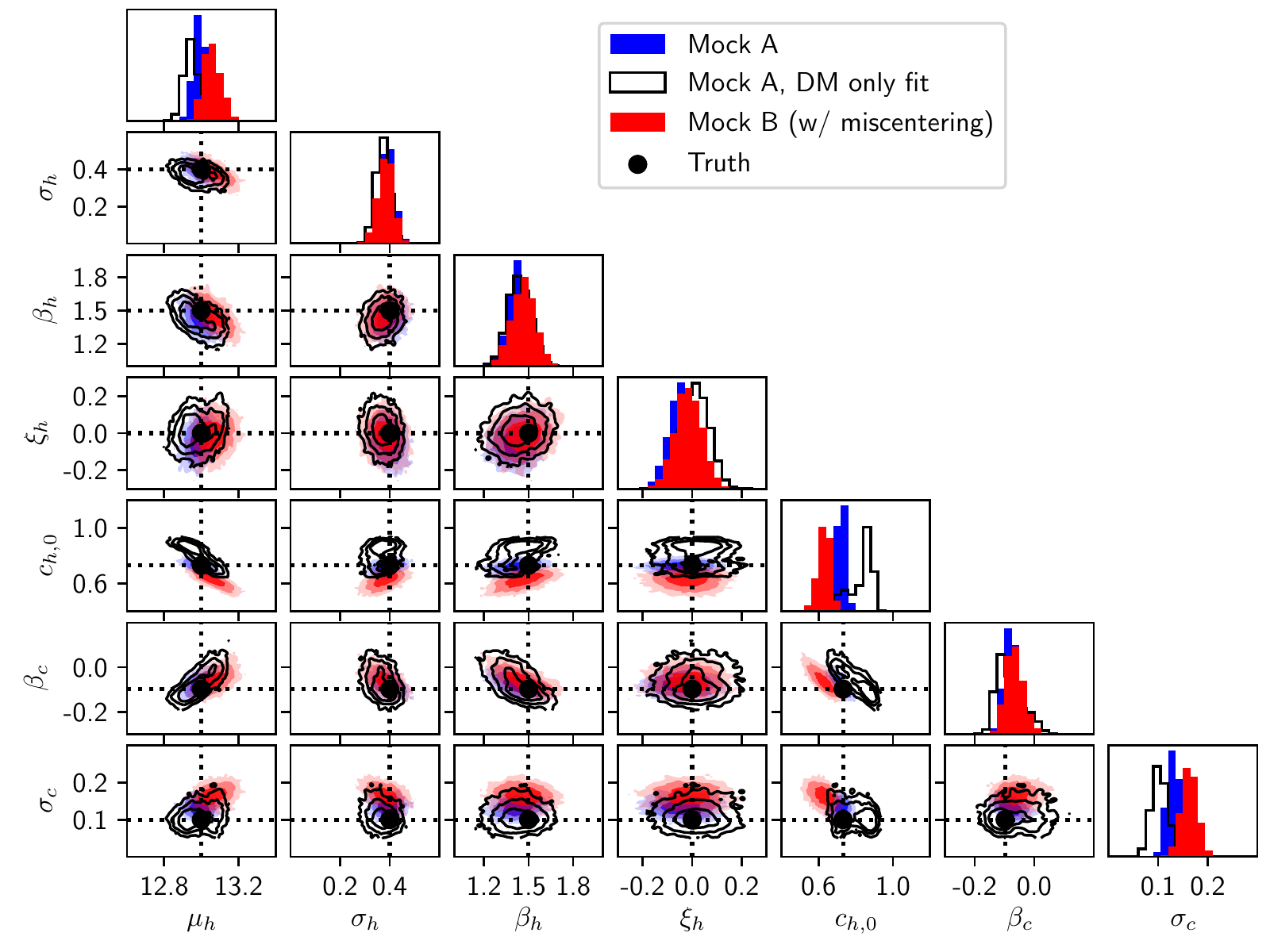}
 \caption{Posterior probability distribution for the hyper-parameters describing the distribution in halo mass and halo concentration, obtained from fits of different models to sets of mock observations A and B. Blue contours show the inference obtained by fitting the model to mock A. Black Solid lines correspond to the inference made on mock A by fitting a model that ignores the contribution of the stars to the lensing signal. Red contours correspond to the inference made by fitting the model to mock B, which differs from mock A by the presence of miscentering between galaxies and halos.
Contours delimit 68\%, 95\% and 99.7\% enclosed probability regions.
Black dots and dotted lines show the true values of the hyper-parameters used to create the mocks.}
 \label{fig:cornerplot}
\end{figure*}

In \Fref{fig:cornerplot} we plot the inferred posterior probability distribution on the hyper-parameters describing the distribution of halo masses and concentrations, along with the true values used to generate the mock.
The inference is accurate. The inference on the coefficient $\xi$ is consistent with zero indicating no correlation between halo mass and stellar mass density at fixed galaxy mass, consistent with the input model. This is perhaps not surprising, since the model we are fitting is identical to the one used to generate the mock. The average halo mass is recovered with a precision of $0.03$~dex.
This value is somewhat arbitrary, as it is essentially set by the size of the mock sample and the number density of sources, which we picked to be 5,000 lens galaxies and $20\,\rm{arcmin}^{-2}$ respectively, but it sets the sensitivity of our experiments to possible systematic effects.

With mock A, we are testing for biases introduced by our treatment of the source redshift uncertainty, which we propagate directly onto the uncertainty in the tangential shear. The fact that we obtain an accurate answer implies that any bias related to this procedure is too small to be detected with our mock (i.e. smaller than $0.03$~dex in halo mass).

\begin{figure}
 \includegraphics[width=\columnwidth]{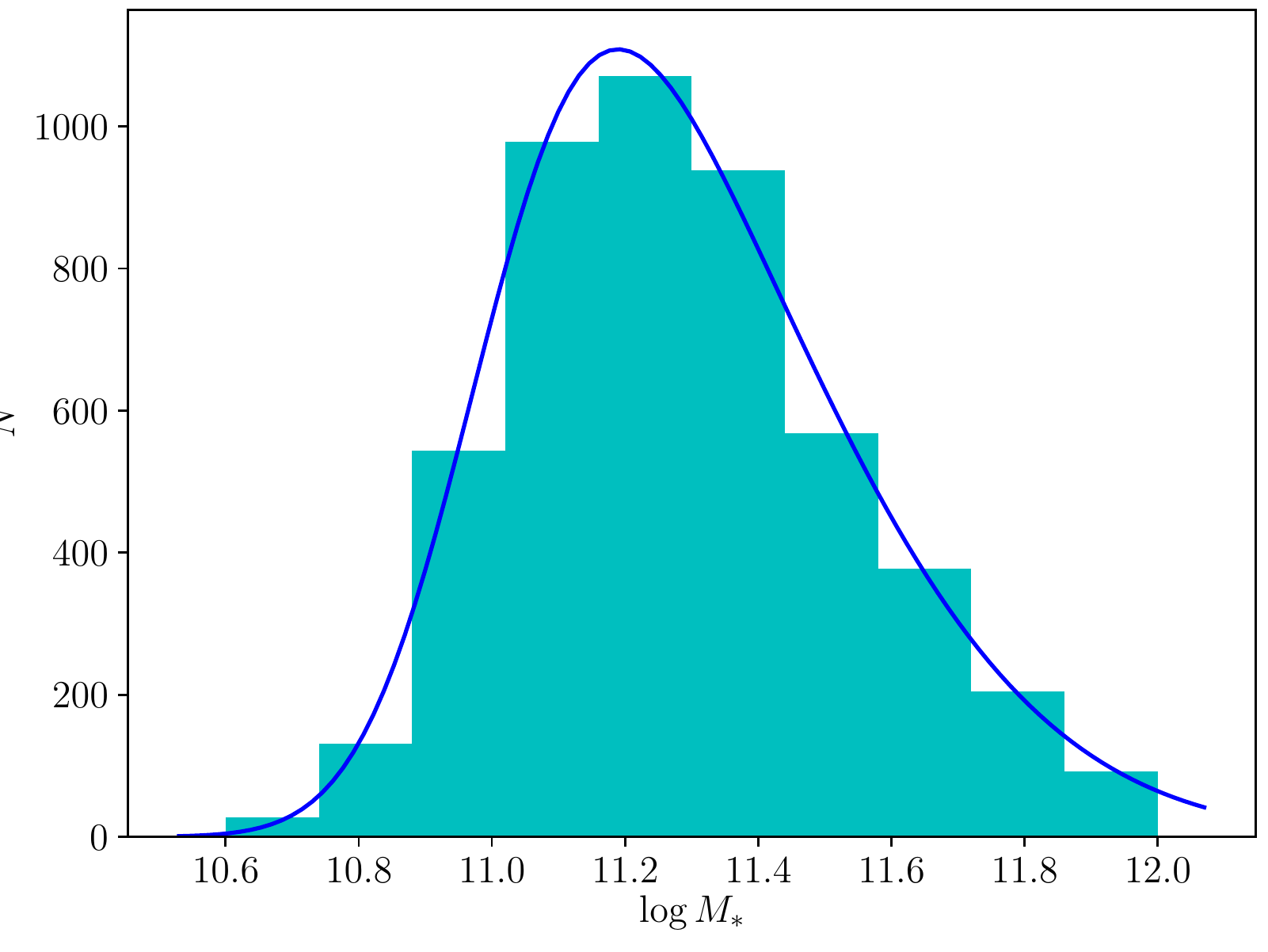}
 \caption{{\em Histogram:} Distribution in stellar mass of the mock A sample. {\em Line:} Maximum likelihood stellar mass distribution, described by \Eref{eq:fullskew}, inferred by fitting the model in subsection \ref{ssec:fullmodel} to the mock data. The corresponding parameter values are $\mu_*=10.99$, $\alpha_*=2.63$, $\sigma_*=0.40$.}
 \label{fig:mstar}
\end{figure}

In addition to the hyper-parameters describing the halo mass distribution, it is also interesting to check whether the model is able to recover the intrinsic distribution in stellar mass of the population.
In our model, the stellar mass distribution is described as a skewed Gaussian, meant to recover the drop at low masses generated by the sharp cut in $\mobs$ applied to define the sample.
In \Fref{fig:mstar} we plot the distribution in true stellar mass of the sample, together with the inferred distribution, as given by the maximum likelihood inference on the skewed Gaussian parameters $\mu_*$, $\alpha_*$, $\sigma_*$.
There is good agreement between the two distributions: the model correctly infers the presence of a tail of galaxies with true mass below the $\log{\mstar}=11$ observational cut. This is a key feature of our method, that allows for an unbiased estimate of the correlation between halo mass and galaxy properties.

\begin{figure}
 \includegraphics[width=\columnwidth]{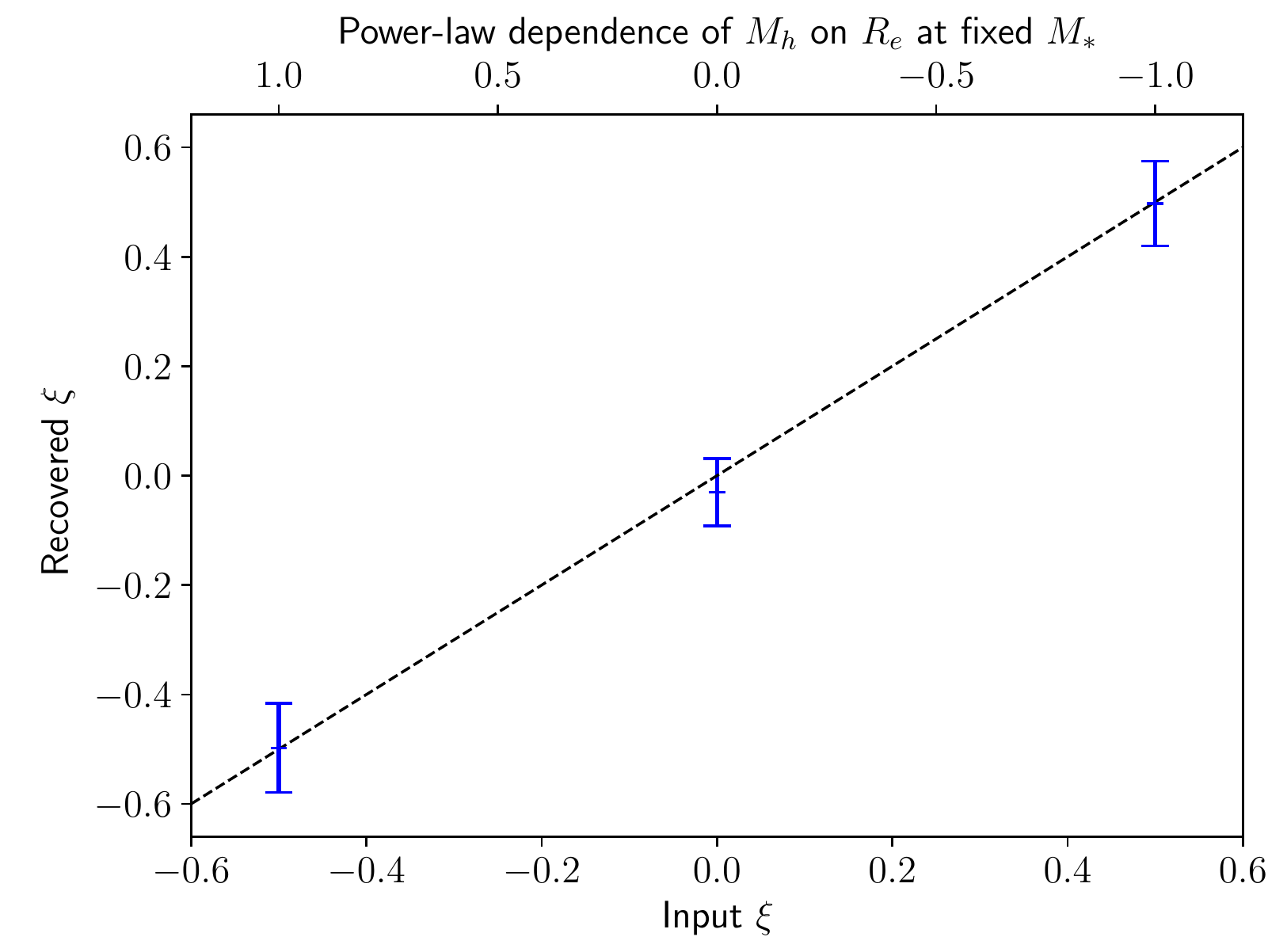}
 \caption{Recovered values of the parameter $\xi$ defined in \Eref{eq:fullmuhalo}, describing a dependence of halo mass on stellar mass density at fixed stellar mass, as a function of the input value, for three different variations of the mock A simulation. The inference is consistent with the input truth in all cases.}
 \label{fig:xipar}
\end{figure}

Mock A was generated assuming that halo masses are independent of galaxy sizes, at fixed stellar mass. Our inference recovers this feature ($\xi=0$). However, we also wish to show that our method can also recover a positive or negative correlation with size, if this is present in the data. 
For this purpose, we generate two new mock populations, similar to mock A, but with an added dependence of halo mass on size: values of $\log{\mhalo}$ are still drawn from a Gaussian distribution with mean given by \Eref{eq:fullmuhalo}, but the value of parameter $\xi$ is set to $-0.5$ and $0.5$.
The two cases correspond to a positive linear dependence of halo mass on galaxy size, at fixed stellar mass, and a negative linear dependence, respectively.
Our model still provides an accurate inference on the parameter $\xi$, as shown in \Fref{fig:xipar}.

\subsection{Mock A: stellar mass contribution to $\Delta\Sigma$}

Before applying our method to a more complex simulation, we can use mock A to study the sensitivity of our inference to the presence of the central galaxy. 
We would like to know what would the inferred halo mass distribution be if we were to neglect the contribution of the stellar mass to the weak lensing signal.
We then fit a simpler version of our model, in which galaxies are treated as massless, to the mock A data. The resulting inference on the hyper-parameters describing the dark matter distribution is plotted in \Fref{fig:cornerplot}.
The inferred average halo mass at $\mpiv$, parameter $\mu_{h,0}$, is consistent with the true value, and does not shift significantly with respect to the case in which the stellar mass is included. However, the inferred average concentration, $\mu_{c,0}$ is overestimated by about $0.1$~dex. This makes sense, as the model compensates the lack of a central component, present in the data, with a more concentrated halo.

\subsection{Mock B: the effect of miscentering}

The second set of simulated weak lensing observations, mock B, is a variation of mock A in which we allow for miscentering between central galaxies and their halos. We do this by adding a random shift in the projected distance between the two mass components, drawn from a Gaussian distribution with dispersion $10\,\rm{kpc}$, as described in subsection \ref{ssec:mockE}.
The inferred hyper-parameters are plotted in \Fref{fig:cornerplot}.
There is only a $0.06$~dex shift in the inference on the average halo mass, moving towards smaller values.
Although the shift would become larger for increasingly higher amplitude of the simulated miscentering, we do not expect large ($>10\,\rm{kpc}$) displacements between central galaxies and halos to be a common occurrence in the real Universe. Therefore we conclude that miscentering has a small impact on studies of this kind.

Miscentering has a somewhat larger impact on the inferred concentration distribution. For instance, the average concentration is underestimated by about $0.10$~dex for this sample.

\subsection{Mock C: a different SHMR}

We now fit our model to a simulation created using a more complex SHMR compared to the simple power-law scaling between stellar and halo mass that we assume in our model.
Mock C is obtained from the more complex mock E, described in \Sref{ssec:mockE}, which is based on the B10 SHMR. 
However, to isolate the effect of the form of the SHMR from other effects included in mock E, we set the lenses at infinite distance from each other. Moreover, we eliminate satellite galaxies from the sample.

In mock E, and therefore also in mock C, dark matter halos are no longer described by a pure NFW profile, but by a smoothly truncated NFW profile, with density profiles described by \Eref{eq:BMO} and truncation radii equal to the virial radius $r_{200c}$.
In the real Universe, the radius at which the dark matter profile drops sharply can in general be different from the virial radius \citep[see][for a discussion]{D+K14}. The truncation radius then introduces an additional parameter, which should in principle be inferred from the data. However, for simplicity, we assume that the ratio between truncation radius and virial radius is known perfectly.
Therefore, the model used to fit this mock, as well as the following mocks, is modified to a smoothly truncated NFW halo with $r_t = r_{200c}$.

\begin{figure*}
 \includegraphics[width=\textwidth]{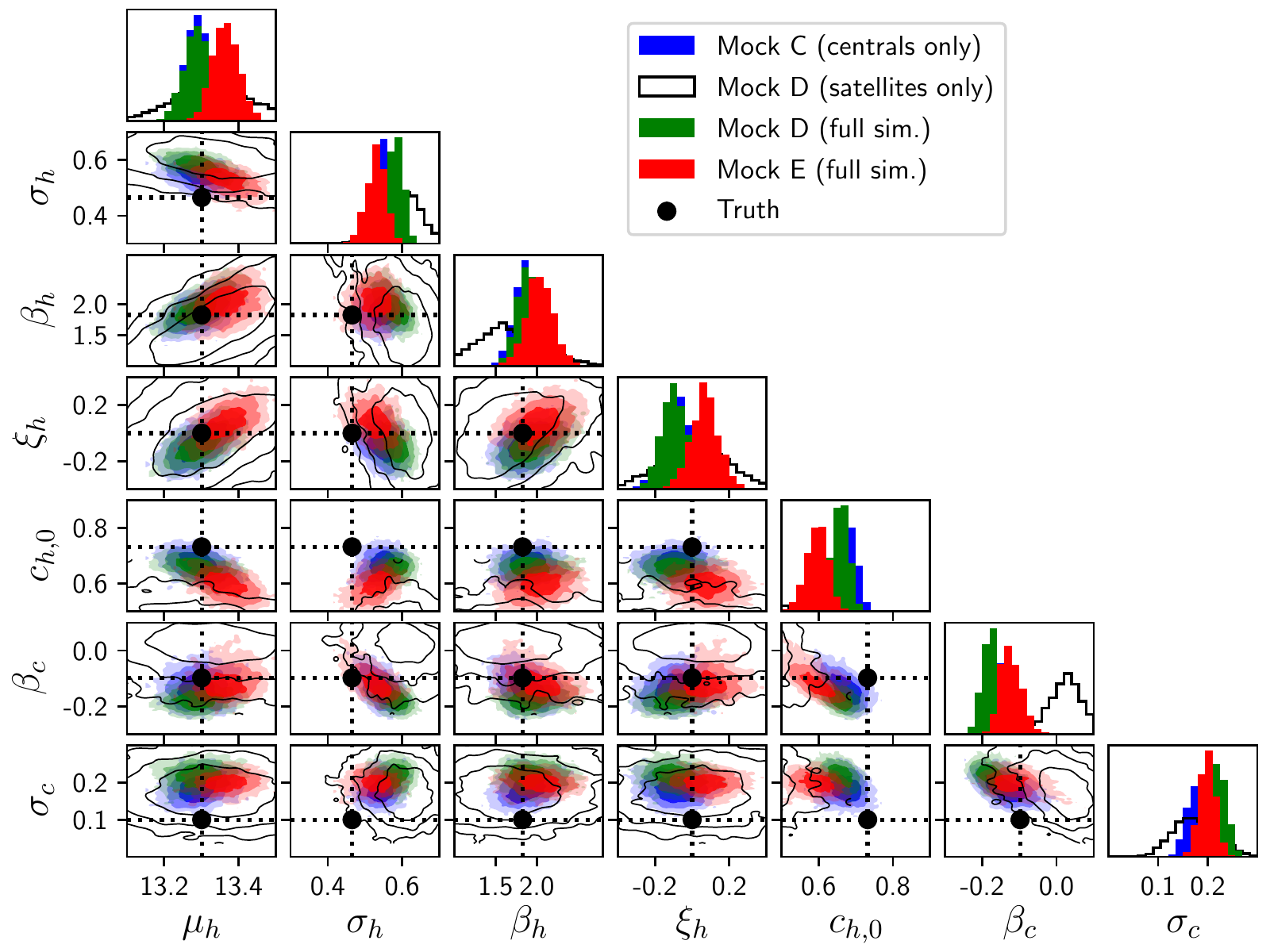}
 \caption{
Posterior probability distribution for the hyper-parameters describing the distribution in halo mass and halo concentration, obtained from fits of different models to sets of mock observations C, D and E. Blue contours show the inference obtained by fitting the model to mock C (a sample with only central galaxies, each at infinite distance from each other). Green contours show the inference obtained by fitting the model to mock D (a sample with both central and satellite galaxies, but no effects from the environment). Red contours correspond to the inference made by fitting the model to mock E (a mock that includes the lensing effect created by neighboring halos).
Contours delimit 68\%, 95\% and 99.7\% enclosed probability regions.
Black dots and dotted lines show the true values of the hyper-parameters used to create the mocks.
}
 \label{fig:mockCDE_cp}
\end{figure*}

\begin{figure*}
 \includegraphics[width=\textwidth]{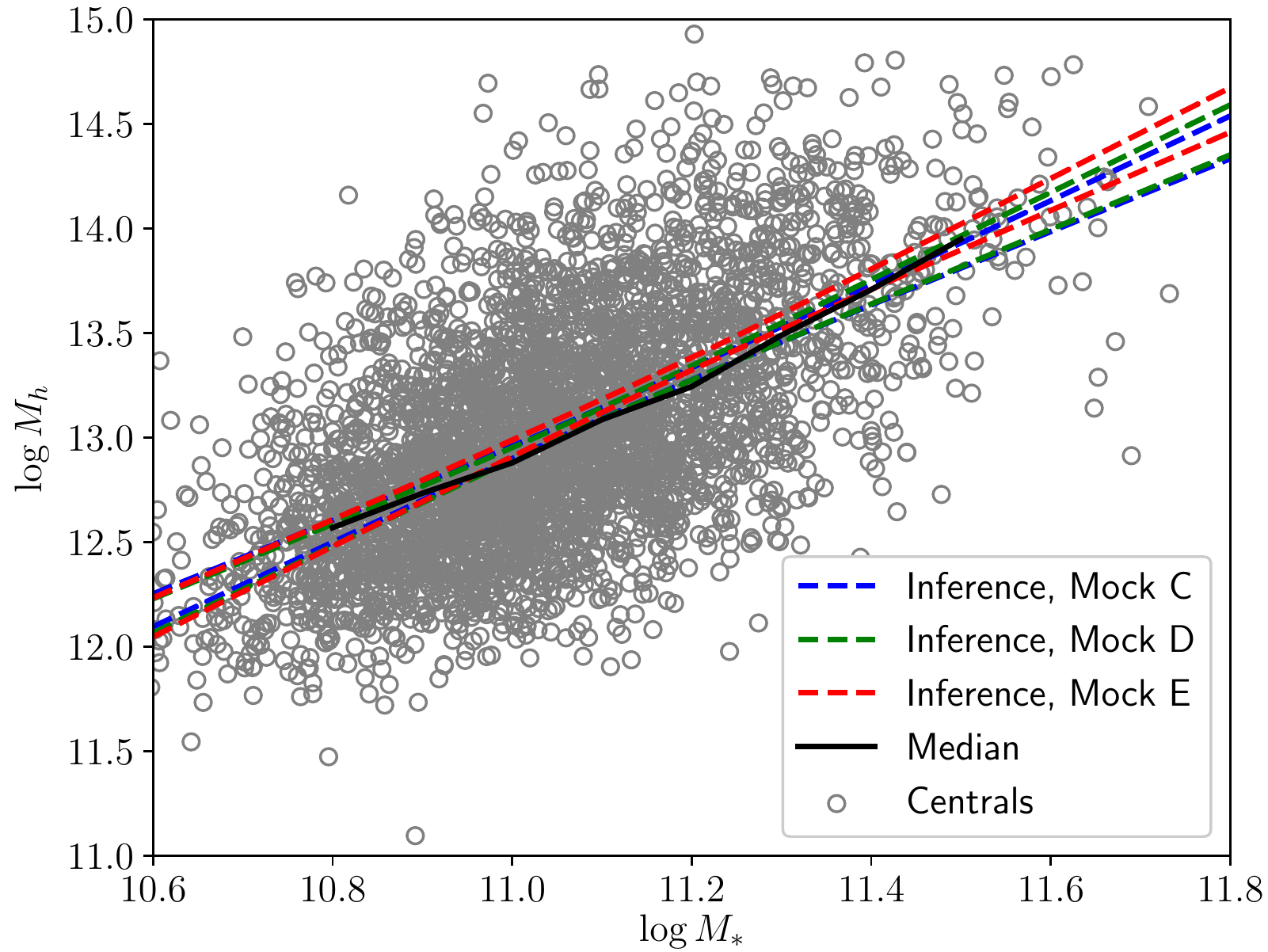}
 \caption{ 
Stellar vs. halo mass for mock galaxies generated with Halotools using the B10 model applied to the halo catalog of the Bolshoi simulation at $z=0$, on which mock C, D and E are based. 
{\em Gray circles}: central galaxies.
{\em Black solid line}: median halo mass as a function of stellar mass, for central galaxies only.
Blue, green and red dashed lines delimit the 68\% confidence region of the inference on the SHMR obtained by fitting our model to mock C, D and E respectively.
}
 \label{fig:shmr}
\end{figure*}

The results are shown in \Fref{fig:mockCDE_cp}. Because the SHMR of this mock is no longer a power-law with constant scatter, there is no obvious definition of the true values of $\mu_{h,0}$, $\sigma_h$ and $\beta_h$ for this sample. We fit a power-law relation to the distribution of $\mhalo$ as a function of $\mstar$, with a minimum least squares method, and use the slope and intercept at $\log{\mpiv}=11.2$ as proxies for the true values of $\mu_{h,0}$ and $\beta_h$.
We then take the standard deviation in $\log{\mhalo}$ around the best fit relation as a proxy for $\sigma_h$.
The true values defined with this procedure are plotted in \Fref{fig:mockCDE_cp} as black dots.
Additionally, we plot in \Fref{fig:shmr} the inferred average halo mass as a function of stellar mass, given by \Eref{eq:fullmuhalo}.

The recovered average halo mass is in very good agreement with the underlying truth, as can be seen from \Fref{fig:mockCDE_cp}. 
Although the true SHMR departs from a pure power-law relation, the inferred model is still a good description of the SHMR, as can be seen by comparing the dashed blue lines in \Fref{fig:shmr}, which delimit the 68\% confidence region of the inference, with the solid black line, corresponding to the median halo mass as a function of stellar mass of the mock.

\begin{figure}
 \includegraphics[width=\columnwidth]{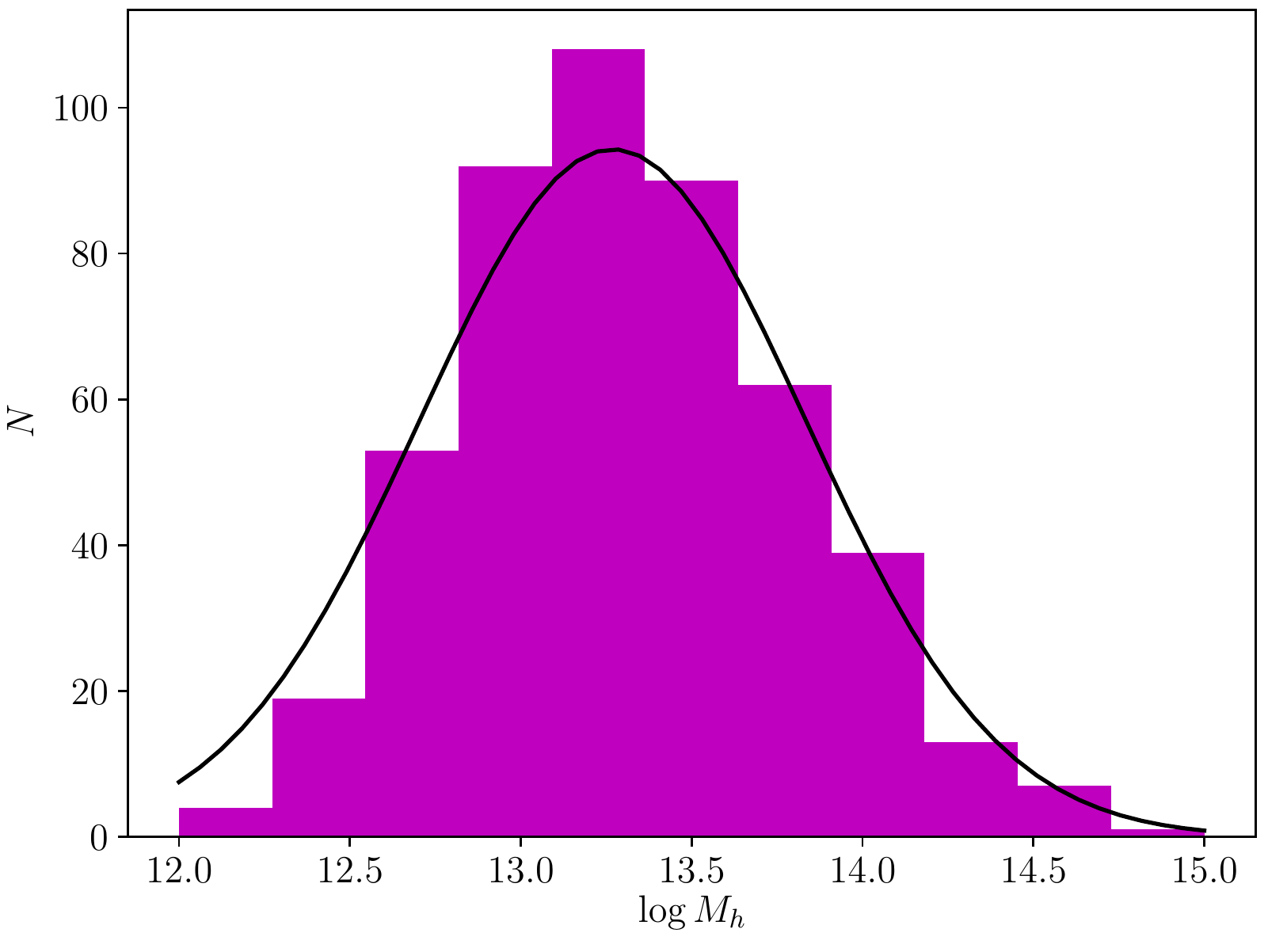}
 \caption{{\em Histogram:} Distribution in dark matter mass of central galaxies in the stellar mass bin $\log{\mstar} \in (11.15, 11.25)$. {\em Line:} Maximum likelihood halo mass distribution at $\log{\mstar}=11.2$, described by \Eref{eq:mhalodist}, inferred by fitting the model in subsection \ref{ssec:fullmodel} to the mock data. The corresponding parameter values are $\mu_{h,0}=13.34$, $\sigma_h=0.46$.}
 \label{fig:mhalodist}
\end{figure}

It is also important to check how the inferred scatter in halo mass compares with that of the mock.
This is shown in 
\Fref{fig:mhalodist}, where we plot the distribution in halo masses for central galaxies with $11.15 < \log{\mstar} < 11.25$, together with the maximum-likelihood inferred distribution at $\log{\mstar}=11.2$.
Our model assumes a Gaussian distribution at fixed stellar mass.
This appears to be a good description of the actual distribution.
The inferred scatter, however, is $\sim0.1$~dex larger than the actual dispersion in halo mass, as can be read from \Fref{fig:mockCDE_cp}.
This is the result of the true SHMR being different from a pure power-law. We expect this small bias to be reduced if we were to use a more accurate model.
As for mock B, the inferred concentration is $\sim0.1$~dex lower than the truth. We understand this to be the result of miscentering.

\subsection{Mock C: measuring the change in slope of the SHMR}\label{ssec:brokenpl}

\begin{figure}
\includegraphics[width=\columnwidth]{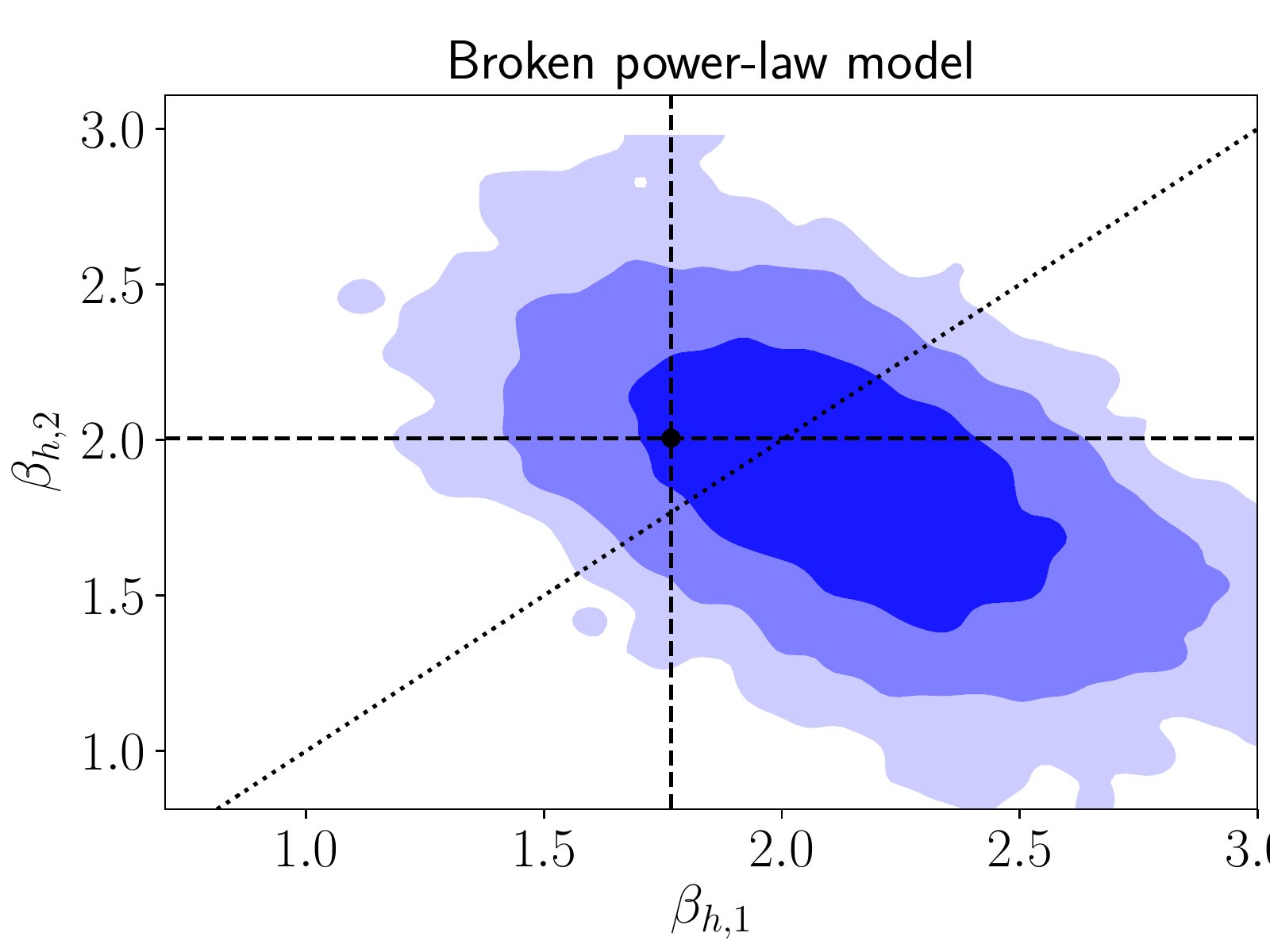}
\caption{Posterior probability distribution obtained by fitting a broken power-law model for the SHMR to the data from the mock C sample, projected on the space defined by the parameters $\beta_{h,1}$ and $\beta_{h,2}$ defined in \Eref{eq:brokenpl}.
The black dot indicates the true values of the parameters, obtained by fitting power-law relations to the distribution of halo mass as a function of stellar mass, above and below $\mpiv$.
The dotted line corresponds to models with a single SHMR slope at all masses: $\beta_{h,1}=\beta_{h,2}$.
}
\label{fig:brokenpl}
\end{figure}

We now fit the data from the mock C sample with a more complex model for the SHMR.
Instead of assuming a power-law relation between the average halo mass and the stellar mass, we allow for an SHMR with a different slope at the low and high mass end, modifying \Eref{eq:fullmuhalo} as follows: 
\begin{multline}\label{eq:brokenpl}
\mu_h(\mstar, \reff) = \mu_{h,0} + \xi_h\log{(\Sigma_*/\Sigma_{*,0})} + \\
\left\{ \begin{array}{ll} \beta_{h,1}(\log{\mstar} - \log{\mpiv}) & \rm{if}\,\,\mstar < \mpiv \\
\beta_{h,2}(\log{\mstar} - \log{\mpiv}) & \rm{if}\,\, \mstar > \mpiv \end{array} \right.
\end{multline}
The equation above corresponds to a broken power-law relation between stellar and halo mass at fixed stellar mass density, with a change in the slope of the SHMR occurring at $\mpiv$.
In principle, $\mpiv$ could be left as a free parameter to be inferred from the data.
For simplicity, however, we keep its value fixed to $\log{\mpiv} = 11.2$. 
In \Fref{fig:brokenpl}, we plot the inference on the parameters $\beta_{h,1}$ and $\beta_{h,2}$, describing the correlation between halo mass and stellar mass below and above $\mpiv$, respectively. The true relation between halo mass and stellar mass is steeper at the high mass end ($\beta_{h,2} > \beta_{h, 1}$). Our broken power-law model recovers this result, although with a large uncertainty. In order to make a more precise inference on the change in slope of the SHMR, a larger sample and/or a larger dynamic range in mass is needed.

Solutions with $\beta_{h,1} = \beta_{h,2}$, corresponding to the pure power-law model considered previously and marked as a dotted line in \Fref{fig:brokenpl}, are within the inferred $1\sigma$ confidence region. This means that models with a broken power-law SHMR do not provide a significantly better description of the data used in our simulation, compared to a single power-law model. Therefore, from here on, we will focus exclusively on the simpler model with a single value of $\beta_h$ over the whole mass range.

\subsection{Mock D: the effect of satellites}

We take one step further in complexity, and add satellite galaxies to the sample.
For these objects, the lensing effect of the main halo, offset from the galaxy, is included.
Lens galaxies are still assumed to be at infinite distance from each other, so that the number of halos distorting sources behind any given galaxy is at most two (one for centrals).
In our model, galaxies are treated as isolated, therefore the presence of a more massive halo offset from it could in principle bias our inference.

We first fit our model to a sample consisting only of satellite galaxies. The inference is shown as solid curves in \Fref{fig:mockCDE_cp}. Since satellites account for only 16\% of the galaxies in the simulation, the uncertainty on the hyper-parameters is broadened due to the smaller sample size.
The main difference with respect to the inference based on mock C, is the value of the mean concentration: satellite galaxies push the parameter $\mu_{c,0}$ towards values as low as allowed by the prior.

We then use the full sample of galaxies, both centrals and satellites. In this case, the inference does not change significantly with respect to the case with centrals only. This can be seen by comparing the green and blue contours and lines in \Fref{fig:mockCDE_cp} and \Fref{fig:shmr}.
We conclude that our method is robust to the presence of a small fraction (16\% in this case) of satellite galaxies, at least in the mass regime probed in our test.

\subsection{Mock E: the full simulation}\label{ssec:full}

We can finally apply our method to the full mock realization of the weak lensing observations, consisting of shape measurements over a contiguous area of 120 square degrees.
This simulation offers an additional challenge with respect to the tests carried out so far.
Up until this point, lenses were at an infinite distance from each other, therefore the assumption of isolated lenses was true by construction.
The model was fit to sources located within an angle $\thetamax$ in projection from each galaxy ($300\,\rm{kpc}$ in physical distance), and there was no ambiguity over which foreground galaxy lensed any given source.

In mock E, and in the real Universe, instead, each source is lensed by every galaxy and halo in its foreground. 
In order to apply the isolated lens approximation, on which our method is based, we must arbitrarily assign background sources to each lens in our sample.
This can be a problem, especially in cases where pairs or multiplets of lens galaxies are found at close projected distances from each other.
We cannot build sets of weak lensing observations within cones of radius $\thetamax$ around every lens galaxy in the sample without using the same data twice.

A possible solution could be using a more complex subdivision of sources among foreground lenses, for example using adaptive boundaries.
Alternatively, we could arbitrarily decide to remove some lens galaxies from the sample until the regions within $\thetamax$ are no longer overlapping.
None of these solutions is ideal.
It would be best to explicitly model the contribution of multiple lenses to each background source. 
This, however, would be computationally very challenging, as it would require, for each draw of a set of hyper-parameters in the MCMC, to marginalize over $3\times N_{lens}$ parameters, where $N_{lens}$ is the number of lenses in the sample, and the factor of $3$ is the number of free parameters for each lens: $\mstar$, $\mhalo$ and $\chalo$. In other words, it would require the evaluation of a $3\times N_{lens}$-dimensional integral. With the isolated lens assumption, the problem reduces to the much more tractable calculation of $N_{lens}$ $3$-dimensional integrals (\Eref{eq:integral2}).

For our test, we adopt the second of the two solutions discussed above: whenever two lens galaxies are located at a projected distance smaller than $2\thetamax$, the least massive of the two, according to the observed stellar mass, is removed from the sample.
The rationale for this choice is that we expect the lensing signal to be dominated by the most massive halo, which we expect to correspond to the most massive galaxy.
With this procedure, we remove 14\% of the lens galaxies.
Many of the excluded galaxies are satellites: the residual satellite fraction is 10\%, from the initial 16\%.

The inferred hyper-parameters are plotted in \Fref{fig:mockCDE_cp}, and the corresponding SHMR is shown in \Fref{fig:shmr}. 
Remarkably, the inference is accurate to better than $0.1$~dex in halo mass over a decade in stellar mass.

\subsection{A biased approach}\label{ssec:bias}

As a last test, we examine what would happen if we were to adopt a more traditional approach to search for a correlation between halo mass and galaxy size at fixed stellar mass, consisting in making a bin in observed stellar mass, splitting it according to size, and measuring a stacked weak lensing signal around galaxies in each bin.

\begin{figure}
 \includegraphics[width=\columnwidth]{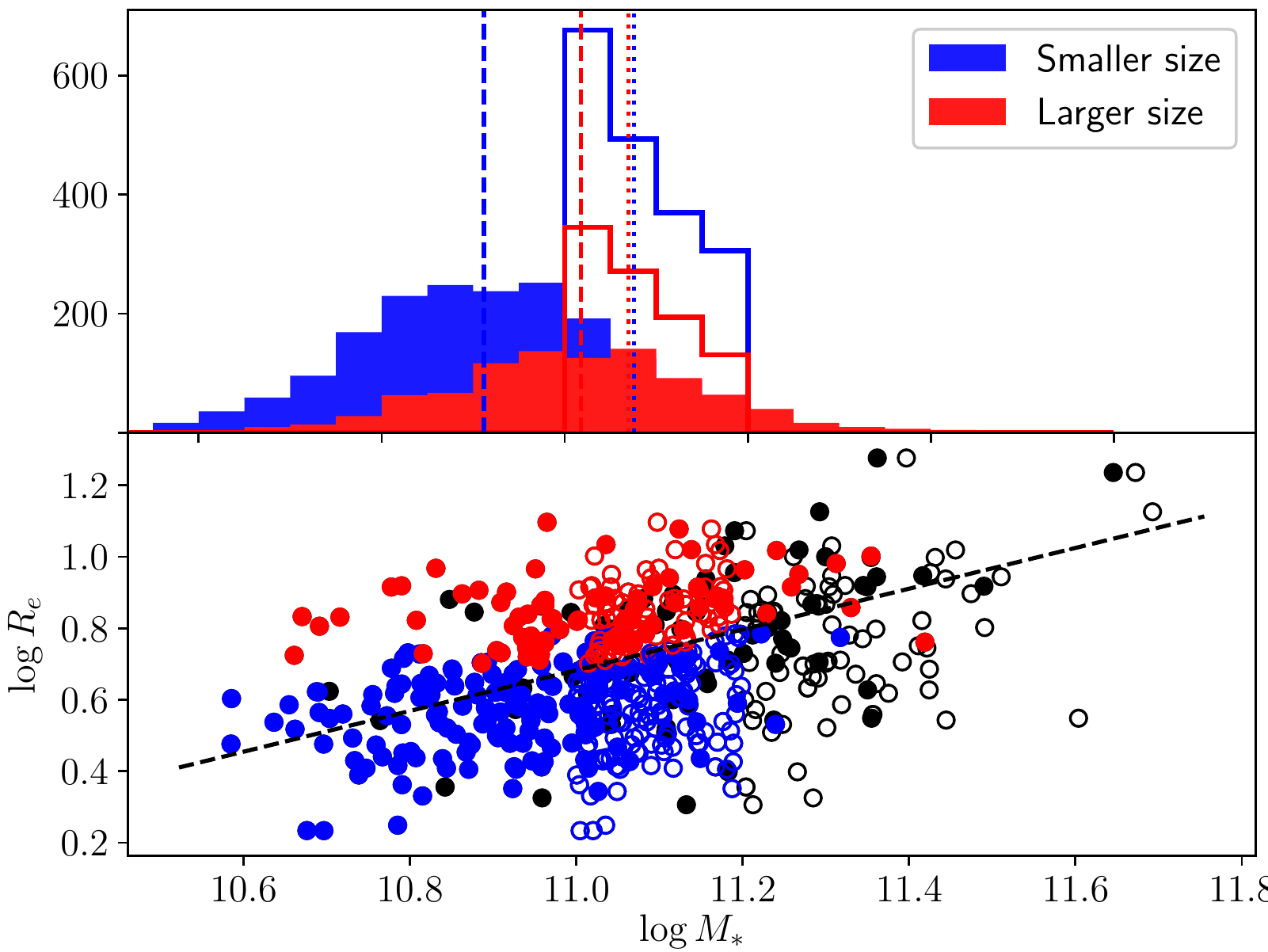}
 \caption{Mass-size relation of mock galaxies. {\em Bottom:} effective radius as a function of stellar mass for a subset of 1,000 objects in the mock sample. Filled circles mark the true stellar mass, while empty circles mark the observed values. Objects in blue (red) correspond to galaxies in the $\log{\mobs}\in[11.0,11.2]$ bin with smaller (larger) sizes compared to the mass-size relation. 
The dashed line shows the mass-size relation used to create the mock.
{\em Top:} Distribution in true stellar mass (filled histogram) and observed stellar mass (empty histogram) of the two subsamples. Vertical dashed (dotted) lines mark the median of the true (observed) stellar mass of each sub-bin.}
 \label{fig:bias}
\end{figure}

Let us take mock E, and make a bin in observed stellar mass with $11.0 < \log{\mobs} < 11.2$.
Let us split this mass bin in two sub-bins: galaxies lying above the mass-size relation at their observed stellar mass go into a `larger size' bin, while galaxies with smaller size compared to the average at their $\mobs$ go into a `smaller size' bin.
In \Tref{tab:bias} we report the mean values of the observed stellar mass, true stellar mass, effective radius and halo mass for the full bin and the two sub-bins.
\begin{table}
 \caption{Mean properties of each $\mobs$ bin and relative sub-bins.}
 \label{tab:bias}
 \begin{tabular}{llccc}
 \hline
$\mobs$ Bin &  & Full bin & Larger & Smaller \\
 \hline
 $[11.0,\,11.2]$ & $\log{M_*^{\mathrm{(obs)}}}$ & $11.08$ & $11.08$ & $11.09$ \\
 & $\log{M_*}$ & $10.98$ & $11.04$ & $10.94$ \\
 & $\log{R_e}$ & $0.66$ & $0.85$ & $0.56$ \\
 & $\log{M_h}$ & $12.81$ & $12.92$ & $12.75$ \\
\hline
$[11.2,\,11.4]$ & $\log{M_*^{\mathrm{(obs)}}}$ & $11.28$ & $11.28$ & $11.28$ \\
 & $\log{M_*}$ & $11.14$ & $11.21$ & $11.11$ \\
 & $\log{R_e}$ & $0.78$ & $0.97$ & $0.67$ \\
 & $\log{M_h}$ & $13.12$ & $13.25$ & $13.05$ \\
\hline
$> 11.4$ & $\log{M_*^{\mathrm{(obs)}}}$ & $11.51$ & $11.50$ & $11.51$ \\
 & $\log{M_*}$ & $11.32$ & $11.39$ & $11.30$ \\
 & $\log{R_e}$ & $0.86$ & $1.08$ & $0.79$ \\
 & $\log{M_h}$ & $13.46$ & $13.71$ & $13.38$ \\
\hline

 \end{tabular}
\end{table}
The two sub-bins have the same mean observed stellar mass. However, their average true stellar mass is different, with the larger size bin containing galaxies that are more massive by $0.06$~dex on average.
The mean halo mass of the two sub-samples is correspondingly different, by an even larger amount: $0.10$~dex.
This is the effect of observational scatter, as we discussed in \Sref{sect:shmr}.
\Fref{fig:bias} shows a graphical representation of the problem.
In the bottom panel, we plot the size of our mock galaxies as a function of their stellar mass. This plot is analogous to that of \Fref{fig:toy}, but obtained with a realistic distribution of masses and sizes.

Two different biases are acting. 
One is the well-known Eddington bias. 
Galaxies get scattered into the mass bin from lower and higher intrinsic masses due to observational errors.
Since the galaxy stellar mass distribution is rapidly declining with increasing mass, there are more galaxies with intrinsically lower stellar mass populating the bin.
This explains why the average true stellar mass of our bins is smaller than the average over the observed values.

The second source of bias is the one described in subsection \ref{ssec:introbias}.
Due to the correlation between mass and size, objects scattering into the bin from higher values of their intrinsic mass have on average larger sizes compared to objects with smaller intrinsic mass.
Therefore, when we split the sample in two size bins, the larger size bin tends to be populated with objects 
with larger intrinsic stellar mass.
Although the amount of the bias in stellar mass between the two sub-bins is relatively small, the corresponding bias in halo mass gets amplified due to the steep relation between stellar and halo mass at the high mass end.
Since the difference in average size between the two sub-samples is $0.3$~dex, and the corresponding difference in the mean halo mass is $0.16$~dex, failure to take this bias into account would lead us to believe that halo mass correlates with the $1/2$ power of $\reff$ at fixed stellar mass.

Note that weak lensing did not enter at all the above argument: this is a general bias that applies to any situation in which one tries to infer separate correlations between a quantity and two different variables that are correlated with each other.

\begin{figure*}
 \includegraphics[width=\textwidth]{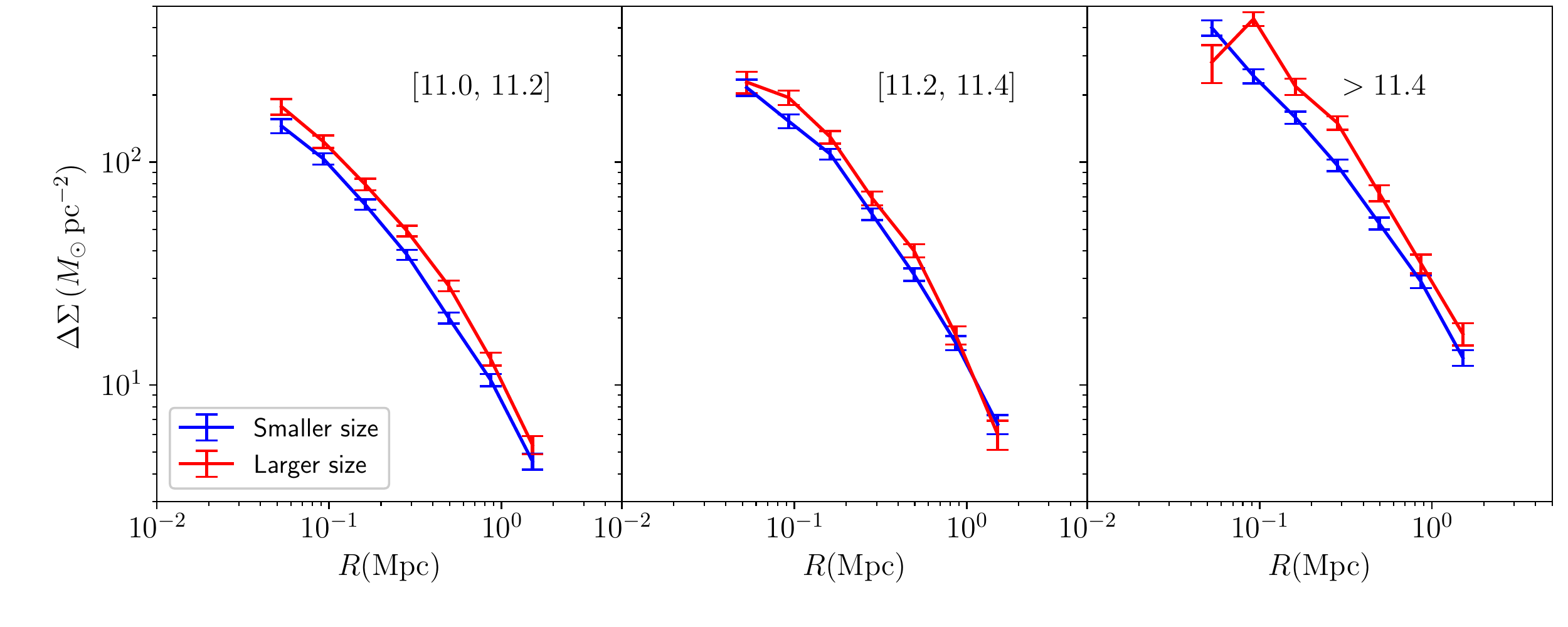}
 \caption{Excess surface mass density in different radial bins around two samples of lens galaxies from mock E, selected by having their observed stellar mass in three different bins (with mass range indicated in each panel), and split according to their position with respect to the mean mass-size relation.
The signal is obtained by stacking weak lensing measurements, using the software {\sc Swot} \citep{Cou++12, SWOT17}. 
In each stellar mass bin, the larger size sample appears to have a higher excess surface mass density compared to the smaller size sample. This is just the result of the two subsamples having different values of their true stellar mass.
\label{fig:stacked}}
\end{figure*}

Let us now obtain a stacked weak lensing signal around galaxies in each sub-bin, and compare them. For this purpose, we use the software {\sc Swot} \citep{Cou++12, SWOT17}. 
We consider two more bins: one covering the range $11.2 < \log{\mobs} < 11.4$, and one with $\log{\mobs} > 11.4$.
The excess surface mass density for the larger and smaller sized galaxies in the three stellar mass bins is plotted in \Fref{fig:stacked}.
In each bin, the signal obtained for the larger size subsample is significantly higher compared to the smaller size subsample.
Since the two curves have been obtained on samples of galaxies of the same observed stellar mass, a naive interpretation of this plot would lead to the conclusion that halo mass correlates with size at fixed stellar mass.
However, such a correlation is not present in the simulation used to produce this plot, therefore this would be a wrong conclusion.

\begin{figure}
 \includegraphics[width=\columnwidth]{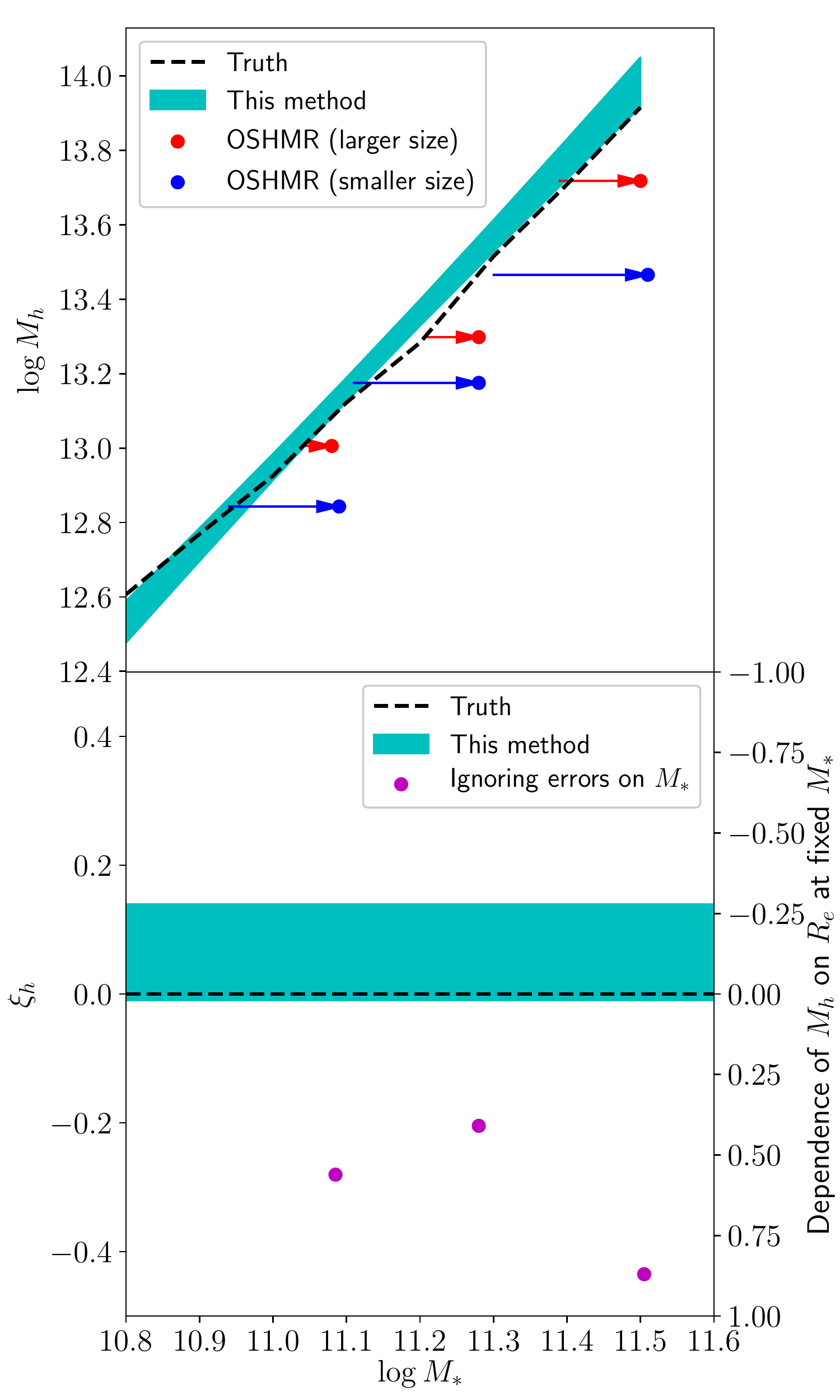}
 \caption{
{\rm Top:} average halo mass as a function of stellar mass of mock E (black dashed line), compared with the values inferred with our method (cyan shaded region). The blue and red circles show the average halo mass of galaxies in each stellar mass bin, as a function of the average value of $\log{\mobs}$ in each bin. The arrows show the effect of Eddington bias on the two subsamples of galaxies. Eddington bias has a stronger effect on the sample of smaller galaxies, leading to an apparent difference in halo mass.
{\rm Bottom:} correlation between halo mass and stellar mass density at fixed stellar mass. The dashed line shows the true value in mock E (no correlation), the cyan band shows the 68\% confidence region of the inference obtained with our method. The magenta dots show the values one would obtain by combining perfect halo mass measurements with the observed values of the average stellar mass and size in each bin.
\label{fig:biased_inference}}
\end{figure}

The effect of Eddington bias, and its differential effect on the two subsamples of galaxies is further illustrated in \Fref{fig:biased_inference}.
In the upper panel, we plot the average halo mass as a function of the mean observed stellar mass in each bin, for large and small galaxies.
This is the OSHMR: the distribution of halo mass as a function of observed stellar mass.
Since the mean observed stellar mass in each bin is larger than the true value, the OSHMR is lower than the truth, for both large and small galaxies.
In addition, large and small galaxies are affected differently from Eddington bias, resulting in a different OSHMR for the two subsamples. 
We stress out that this definition of large or small is only in relation to the {\em observed} stellar mass of each galaxy.

In the bottom panel of \Fref{fig:biased_inference}, we plot the values of $\xi_h$ (the correlation between halo mass and stellar mass density at fixed stellar mass) one would infer by interpreting the difference in OSHMR between larger and smaller galaxies as real.
As expected, the result is biased.
Our method, on the other hand, is able to recover the correct answer.

\section{Discussion and summary}\label{sect:discuss}


We introduced a Bayesian hierarchical inference method to infer the SHMR of a population of massive galaxies, generalized by allowing for a secondary dependence of halo mass on stellar mass density.
Our method is similar in spirit to existing maximum-likelihood approaches, but it differs from them in two fundamental aspects.
Firstly, it explicitly models the intrinsic scatter in the SHMR.
The second, more subtle, difference, is that it makes a clear distinction between true and observed quantities. As shown in \Sref{sect:shmr}, making such a distinction and carefully modeling the effects of observational uncertainties is crucial in order to make accurate inferences.

We tested the method on mock observations generated with semi-analytic models with increasing degrees of complexity, and showed that it can recover accurate halo masses to within $\sim0.1$~dex over a decade in stellar mass.
Although our mocks are somewhat simplistic in their nature (we have not allowed for departures from spherical symmetry in the mass distribution of our galaxies, for instance), our tests have allowed us to gauge the role of two important systematic effects: the role of satellites and that of the environment, which are shown to be small in the mass regime considered ($\log{\mstar} \gtrsim 11$).

Our model is also able to infer the mass-concentration relation of the halos.
Although miscentering between galaxies and their halos can introduce a $\sim20\%$ bias on the average concentration, the correlation between concentration and halo mass appears to be robust with respect to the potential sources of systematic uncertainty explored in our work.
This opens up interesting possibilities for the exploration of correlation between concentration and other galaxy properties at fixed halo mass.
The concentration of a halo is believed to be tightly linked to its formation time \citep{NFW97, Wec++02, Zha++03}, and is in principle sensitive to baryonic physics effects, such as adiabatic contraction.
Measuring correlations between galaxy properties and halo mass then has ample potential for discovery. This is one of the possible uses of our inference method.

Nevertheless, we point out that, like all methods based purely on weak lensing data, our model suffers from the mass-sheet degeneracy, which sets a fundamental limit to the ability to robustly determine the lens density profile. With our approach, we are artificially breaking the degeneracy by asserting a specific form for the density profile of the lenses. However, in order to robustly break this degeneracy, complementary information, such as magnification measurements, is needed. 

Our method offers various advantages over more traditional approaches, such as stacking and maximum-likelihood methods.
It allows for the unbiased exploration of secondary dependences of halo mass on quantities other than stellar mass, such as size, as shown in \Sref{sect:results}.
The same is not true if a simple stacked weak lensing method is used, as our test in subsection \ref{ssec:bias} clearly shows.
Recently, \citet{Cha++17} claimed a detection of a positive correlation between halo mass and galaxy size at fixed stellar mass, $M_h \propto \reff ^ {\eta}$, with $\eta = 0.42\pm0.12$, obtained by comparing the stacked weak lensing signal around galaxies samples split by size at fixed observed stellar mass.
It is possible that this correlation might in part be the result of the bias described qualitatively in subsection \ref{ssec:introbias} and, quantitatively, in \ref{ssec:bias}. In fact, the value of the correlation measured by \citet{Cha++17} is very similar to the value found in our simulation when ignoring the effects of observational scatter in $\mstar$ (see bottom panel of \Fref{fig:biased_inference}).

Stacked weak lensing studies of the correlation between halo mass and galaxy colour at fixed stellar mass are also susceptible to the same bias.
Red galaxies are on average more massive than blue galaxies: for instance, only a small fraction of galaxies at $\log{\mobs} > 11$ are blue. This means that, when binning in observed stellar mass, Eddington bias affects red and blue galaxies differently: for the same value of $\mobs$, red galaxies have, on average, an intrinsically larger stellar mass compared to blue galaxies.
Therefore, the measured difference in the average halo mass between red and blue galaxies at fixed observed stellar mass \citep{Man++16} might also be, in part, the result of the bias described in subsection \ref{ssec:introbias}.
\citet{MNW17} argued that the observed difference 
is a result of scatter. However, they did not specify whether they referred to intrinsic or observational scatter. As we explained in \Sref{sect:shmr}, this distinction is instead very important, as it leads to drastically different physical interpretations.

Another advantage of our method, although not exploited in this work, is that it can be extended to allow for the inclusion of datasets other than weak lensing, such as X-ray emission from the diffuse halo gas, or galaxy kinematics.
This is simply done by adding multiplicative terms to the likelihood, as long as it is possible to predict the corresponding observables from the model.

The method is based on a crucial approximation: the assumption that all lenses are isolated, or, in other words, that for a given background source there is only one lens contributing to its lensing distortion.
Although this approximation breaks down for satellite galaxies and for pairs of halos in close proximity, our tests on mocks show that we can still accurately recover the SHMR, if we restrict the analysis to lens galaxies more massive than $\sim10^{11}M_\odot$, and sources located within $300\,\rm{kpc}$ in projection.

We stress out that the necessity for the isolated lens assumption is purely technical, since it arises from the difficulty of exploring the highly-dimensional parameter space of the full problem, in which each source is lensed by every lens in the sample.
We do not exclude the possibility that this difficulty could be overcome with the use of more sophisticated sampling techniques.

A possible step forward can be made, with the tools already at hand, by abandoning the isolated lens assumption in favor of a less strong {\em isolated group} assumption. When a few galaxies are found to lie in close projected distance from each other, it is possible to explore the parameter space of all the lenses in the ``group'' (which does not need to correspond to a physically bound association) self-consistently, in a finite computational time.
This could, on the one hand, reduce the systematic uncertainties due to neglecting the contribution of neighboring halos to the lensing signal around each group galaxy, and, on the other hand, would get rid of the need to eliminate lenses from the sample in case of overlapping regions of influence.
We leave the exploration of such models for future work.
At the moment, our method is most suited to samples of massive ($\log{\mstar} \gtrsim 11$) galaxies, probing the group and cluster regime, for which the satellite fraction is small and galaxies are well separated in projected distance.


\section*{acknowledgments}
We thank Surhud More and Jiaxin Han for helpful discussion.
This work was supported by World Premier International Research Center Initiative (WPI Initiative), MEXT, Japan, and the National Science Foundation under Grant No. NSF PHY17-48958.
AS is partly supported by KAKENHI Grant Number JP17K14250. 
\bibliographystyle{mnras}
\bibliography{references}

\end{document}